\documentclass[reprint,superscriptaddress,preprintnumbers,amsmath,amssymb,aps,prd,nofootinbib]{revtex4-1}

\usepackage{placeins} 
\usepackage[normalem]{ulem} 
\usepackage{graphicx}  
\usepackage{xcolor}
\usepackage{dcolumn}   
\usepackage{bm}        
\usepackage{hyperref}  
\usepackage{multirow}
\usepackage{siunitx}
\graphicspath{
  {figures/}
}

\def\be{\beta}

\def\ep{\epsilon}

\def\ka{\kappa}
\def\la{\lambda}

\def\pa{\partial}
\def\half{\frac{1}{2}}

\def\bB{{\mathbf{B}}}
\def\bD{{\mathbf{D}}}

\def\mcL{{\mathcal L}}

\newcommand{\ben}{\begin{equation}}
\newcommand{\een}{\end{equation}}
\newcommand{\bea}{\begin{eqnarray}}
\newcommand{\eea}{\end{eqnarray}}
\newcommand{\ba}{\begin{array}}
\newcommand{\ea}{\end{array}}
\newcommand{\bit}{\begin{itemize}}
\newcommand{\eit}{\end{itemize}}

\newcommand{\vev}[1]{\left\langle#1\right\rangle}

\newcommand{\cl}{l_{\phi}}
\newcommand{\bFT}{\beta_\text{ft}}
\newcommand{\cs}{c_\text{s}}

\newcommand{\dx}{\delta x}
\newcommand{\Df}{\Delta f}
\newcommand{\fNG}{f_\text{NG}}

\newcommand{\ie}{\textit{i.e.\ }}
\newcommand{\gmu}{G\mu}

\newcommand{\Llag}{\ell_\mcL}
\newcommand{\vLag}{\bar v_\mcL}
\newcommand{\LlagIni}{\ell_{\mcL,\text{init}}}
\newcommand{\Ltilde}{\tilde\mcL}

\newcommand{\ellInit}{\ell_\text{init}}

\newcommand{\SwidG}{w_\text{g}} 
\newcommand{\SwidS}{w_\text{s}} 

\newcommand{\tStart}{t_{\rm start}}

\newcommand{\tDiff}{t_{\rm diff}}

\newcommand{\tLife}{t_\text{life}}
\newcommand{\tIni}{t_\text{init}}
\newcommand{\tEva}{t_\text{eva}}
\newcommand{\vAv}{\bar{v}}

\begin{document}

\preprint{HIP-2021-15/TH}

\title{Loop decay in Abelian-Higgs string networks}

\newcommand{\Sussex}{\affiliation{
Department of Physics and Astronomy,
University of Sussex, Falmer, Brighton BN1 9QH,
U.K.}}

\newcommand{\HIPetc}{\affiliation{
Department of Physics and Helsinki Institute of Physics,
PL 64, 
FI-00014 University of Helsinki,
Finland
}}

\newcommand{\EHU}{\affiliation{
Department of Physics,
University of the Basque Country UPV/EHU, 
48080 Bilbao,
Spain
}}

\author{Mark Hindmarsh}
\email{mark.hindmarsh@helsinki.fi}
\HIPetc
\Sussex

\author{Joanes Lizarraga}
\email{joanes.lizarraga@ehu.eus}
\EHU

\author{Ander Urio}
\email{ander.urio@ehu.eus}
\EHU

\author{Jon Urrestilla}
\email{jon.urrestilla@ehu.eus}
\EHU

\date{\today}

\begin{abstract}
We study the decay of cosmic string loops in the Abelian-Higgs model. We confirm 
earlier results that loops formed by intersections of infinite strings formed from random-field initial conditions 
disappear quickly, with lifetime proportional to their initial rest-frame length $\ellInit$. 
We study a population with $\ellInit$ up to $6000$ inverse mass units, and measure the 
proportionality constant to be $0.14\pm0.04$, independently of the initial lengths.
We propose a new method to construct oscillating non-self intersecting loops 
 from initially stationary strings, and show that by contrast these loops have lifetimes scaling approximately as $\ellInit^2$,
in line with previous works on artificially created string configurations.  
 We show that the oscillating strings have mean-square velocity $\vAv^2 \simeq 0.500 \pm 0.004$, consistent with the 
Nambu-Goto value of $1/2$, while the network loops have $\vAv^2 \simeq 0.40 \pm 0.04$.
We argue that whatever the mechanism behind the network loop decay is, it is non-linear, 
can only be suppressed by careful tuning of initial conditions, and is much stronger than gravitational radiation. 
An implication is that one cannot use the Nambu-Goto model to derive 
robust constraints on the tension of field theory strings. 
We advocate parametrising the uncertainty 
as the fraction $\fNG$ of Nambu-Goto-like loops surviving to radiate gravitationally. 
None of the 31 large network loops created survived longer than 0.25 of their initial length, 
so one can estimate  that $\fNG<0.1$ at $95$\% confidence level. 
If the recently reported NANOgrav signal is due to cosmic strings, 
$\fNG$ must be greater than $10^{-3}$ in order not to violate bounds from the 
Cosmic Microwave Background.
\end{abstract}

\maketitle

\section{Introduction}

In the traditional picture of cosmic string evolution \cite{VilShe94,Hindmarsh:1994re}, 
one models the strings as idealised line-like objects, 
which are expected to evolve according to the Nambu-Goto equation \cite{Forster:1974ga,Arodz:1995dg,Anderson_1997}, 
with an additional reconnection rule when strings cross \cite{Shellard:1987bv,Matzner:1988qqj,Achucarro:2006es}. 
In this model, 
the most stringent constraints are provided by the generation of a background of gravitational waves (GWs) 
by slowly-decaying loops of cosmic string, 
currently $\gmu \lesssim 10^{-10}$ \cite{Abbott:2017mem,Ringeval:2017eww,Blanco-Pillado:2017rnf,Auclair:2019wcv,Auclair:2019jip} 
(with $\mu$ the string tension and $G$ Newton's constant).
Indeed, the recent observations of an excess in the timing residuals of millisecond pulsars \cite{Arzoumanian:2020vkk} 
could be accounted for by GWs from Nambu-Goto strings with tension saturating the bound \cite{Blasi:2020mfx,Ellis:2020ena}.

On the other hand, simulations of string networks based on an underlying classical Abelian-Higgs (AH) field theory 
\cite{Vincent:1997cx,Moore:2001px,Bevis:2006mj,Bevis:2010gj,Daverio:2015nva,Correia:2019bdl,Correia:2020gkj,Correia:2020yqg}
show that the most important decay channel is the production of classical scalar and gauge radiation, in 
which case the strongest constraints come from a chain of decays into $\gamma$-rays. 
The observed diffuse $\gamma$-ray background then limits the 
string tension to be $\gmu < 2.7 \times 10^{-11} \bFT^{-2}$, where $\bFT^2$ is the branching fraction of the scalar and gauge decays into 
Standard Model particles (most likely the Higgs, through a portal coupling) \cite{Mota:2014uka}.

In either case, the amplitude of Cosmic Microwave Background (CMB) fluctuations bounds the string tension as $\gmu \lesssim 10^{-7}$ \cite{Lazanu:2014eya,Lizarraga:2016onn,Ade:2013xla}.

The discrepancy is a consequence of the qualitative and quantitative differences between the descriptions of both models. 
In the NG model loops of string are assumed to decay only by gravitational radiation, which gives them a lifetime $\tLife \sim (\gmu)^{-1} \ellInit$, 
and considerably enhances the gravitational signal.
On the other hand, in AH simulations, where the decay channels of the field theory are not neglected, loops of initial length  $\ellInit$ disappear in a time $\tLife < \ellInit$  
\cite{Vincent:1997cx,Hindmarsh:2008dw}.

The mechanism behind the rapid decay of AH string loops is still not understood. 
Initial work on string decay focused on the quantum production of particles by the classical fields 
\cite{Vachaspati:1984yi,Srednicki:1986xg,Brandenberger:1986vj}, 
which is negligible, and in any case not relevant for purely classical simulations. 
Numerical simulations studied sinusoidal standing waves  \cite{Olum:1998ag} 
and collisions of travelling waves leading to cusps \cite{Olum:1999sg} 
on strings wrapping the simulation volume. 
For sinusoidal standing waves, the massive radiation power decreases exponentially with the ratio 
of the string curvature to the string width  \cite{Olum:1998ag}, as expected by linear 
perturbation theory, demonstrating that this is not 
the explanation of the energy loss in collapsing loops.\footnote{A claim of massive radiation power from a sinusoidal standing wave decreasing as a power law with the curvature to width ratio \cite{Vincent:1997cx} could be explained by perturbations in the initial field configuration.}

Recently a study of the decay of non-self-intersecting 
loops in the Abelian Higgs model was presented \cite{Matsunami:2019fss}.  
The loops were created by ingenious initial conditions. which made them spin and oscillate 
on trajectories which are presumably well approximated by the Nambu-Goto equations.  
It was shown that the lifetime of these loops scaled as $\ellInit^2$. 
A model of energy loss in terms of radiation from kink collisions on a Nambu-Goto string was constructed. 
Combining this $\ellInit^2$ decay behaviour with the standard NG scenario in which loops radiate GWs at a constant rate, 
it was argued that the model supported the NG description of field theory strings at large times, when  
GW radiation eventually dominates over scalar and gauge radiation. 

On the other hand, there has been no indication of such long-lived loops in AH network simulations to date, which would show up as a 
slowing of the decay of string length, and a departure from scaling. Scaling is generally quantified as a 
linear growth in the mean string separation, which is very closely followed in the largest  
numerical simulations to date \cite{Hindmarsh:2017qff,Correia:2020gkj}, where the curvature to width ratio is many hundreds. 
However, even in these simulations there 
are few loops as large as the largest of those studied in Ref.~\cite{Matsunami:2019fss}, which were of order $10^3$
inverse mass units in length, and those that are this large are not followed until they disappear. 
So it is conceivable that the difference between the ``artificial'' loops and the network loops is one of size. 

In this paper we examine this possibility, by studying loops created from the same random initial conditions 
used in cosmological network simulations \cite{Hindmarsh:2017qff}. 
We take the spacetime to be Minkowski for simplicity, since 
the issues we address, those of energy loss and stability of solutions, 
appear in all homogeneous backgrounds. 
The loops are of similar and larger sizes to those in Ref.~\cite{Matsunami:2019fss}. We  
follow their evolution until they disappear, and record their lifetimes, 
finding no evidence for long-lived loops in this population. We confirm the rapid energy loss observed in 
Ref.~\cite{Hindmarsh:2008dw}, and establish the 
linear relation between lifetime $\tLife$ and initial loop length $\ellInit$ more quantitively, as 
$\tLife = (0.14 \pm 0.04)\ellInit$. 
This decay mechanism is clearly much stronger than gravitational radiation. 

We also study artificial loops in field theory, created by methods similar to those of Ref.~\cite{Matsunami:2019fss}. 
We find that, after an initial transient period of energy loss, 
the rest frame length $\ell$ of this population of loops decays much more slowly, with a lifetime scaling as $\ellInit^2$, 
as in Ref.~\cite{Matsunami:2019fss}.  

The mean square string velocities of both types of loops is also measured, using the 
estimators developed in Ref.~\cite{Hindmarsh:2017qff}. Artificial loops have mean square velocity $0.5$, as expected by Nambu-Goto dynamics in Minkowski spacetime.  
On the other hand,  the mean square velocity of network loops has a wider range of values around $0.4$, 
lower than the Nambu-Goto prediction.

Our results confirm that it is possible to create field configurations whose evolution is well approximated 
by Nambu-Goto dynamics in the limit of small curvature, which was perhaps not in doubt. 
However, our results also confirm that the loops created by network evolution are typically not in this category, 
even at very low curvatures where the Nambu-Goto approximation is traditionally expected to apply. 
Network loops decay more or less as quickly as allowed by causality independent of their size, and 
their lower velocity is also not accounted for in the Nambu-Goto approximation.

The rapid size-independent decay points towards a non-linear mechanism, which can transport energy over a large range of length scales.  
We outline a model of loop decay involving interacting massive degrees of freedom on the string, which 
can account for both the behaviour of the Nambu-Goto-like strings whose lifetime scales as $\ellInit^2$, 
and the network loops whose lifetime scales as $\ellInit$. 
Loops decay slowly if the modes are a small perturbation, and 
 rapidly if they contribute significantly to the energy per unit length of the string. 
It remains unclear how the modes are excited. Some kind of instability is 
present, which is perhaps no surprise in a non-linear field theory, but 
understanding it requires further study. 

Without a better understanding, we remain uncertain what are the important decay channels of a string network.
There is no support from our simulations for the Nambu-Goto approximation applying to network loops, which means that the traditional picture of 
gravitational wave production by an assumed population of  long-lived oscillating loops cannot be used to 
derive robust constraints on the tension of field theory strings. 

The best that can be done with our current state of knowledge is to parametrise the uncertainty 
by allowing a fraction $\fNG$ of loops to behave like the artificial loops and survive to radiate gravitationally. 
As we have found no long-lived loops out of the 31 network loops generated by our random initial conditions, 
one might infer that this fraction is bounded above by $0.1$ (at  95\% confidence level),  assuming our loops are a fair sample.  
A dedicated study with much larger statistics is required to obtain firmer limits. 


\section{Model and estimators}
\label{sec:ModelSims}

We study the Abelian Higgs model, as the simplest gauge field 
theory with strings. 
The Lagrangian density for the Abelian Higgs model in Minkowski spacetime is
\ben
L= \Big(D_\mu\phi^*D^\mu\phi + V(\phi) + \frac{1}{4e^2}F_{\mu\nu}F^{\mu\nu}\Big) 
\label{lagrangian},
\een
where $\phi(x)$ is a complex {scalar} field,  
$A_\mu(x)$ is a {vector} field, 
the covariant derivative is \(D_\mu = \partial_\mu - iA_\mu\), 
and the potential is \(V(\phi) = \frac{\la}{4}(|\phi|^2 - \phi_0^2)^2\).

The resulting field equations in the temporal gauge ($A_0 = 0$) are 
\bea
\ddot\phi -D_j D_j \phi + \frac{\lambda}{2} (|\phi|^2 -\phi_0^2)\phi &=& 0, \\
\pa^\mu F_{\mu\nu}  -  e^2{\rm Im}(\phi^*D_\nu\phi) &=& 0.
\label{eom}
\eea
The equations have static cylindrically symmetric solutions, Nielsen-Olesen (NO) vortices \cite{Nielsen:1973cs}. 
The physical length scales in the solution are the Compton wavelengths of the scalar and gauge fields,
\ben
\label{e:SwidDef}
\SwidS = (\sqrt{\la}\phi_0)^{-1}, \quad
\SwidG = (\sqrt{2}e\phi_0)^{-1},
\een
which set the scales on which the fields exponentially approach the vacuum. All our simulations are performed 
with $\la = 2$ and $e=1$ so that the length scales are equal.

We solve the partial differential equations (\ref{eom}) numerically, using the prescription 
presented \cite{Hindmarsh:2017qff},  in cubic lattices, with periodic boundary conditions. 
We refer the reader to that publication for details of the discretisation procedure,
noting here that in preparing the initial conditions, we use a period of diffusive evolution according to the equations 
\bea
\dot{\phi} &=& D_jD_j\phi - \frac{\lambda}{2}(|\phi|^2 - \phi_0^2)\phi \, ,\nonumber\\
F_{0j} &=& \partial_iF_{ij} - e^2 {\rm Im} (\phi^*D_j\phi) \, .
\label{AHdiff}
\eea

\subsection{Length and velocity estimators}

Our main diagnostics for string dynamics are 
positions and velocities extracted from the field configurations. 
We will use different approaches in order to do so.

We can identify the position of strings 
by scanning the lattice and computing the winding number in each plaquette \cite{Kajantie:1998bg}. Those plaquettes with non-trivial winding number are the ones pierced by strings. We can estimate the total length of the string by counting the plaquettes pierced by a string, multiplying that number  by the lattice spacing $\delta x$, and correcting it by a factor of $2/3$ to account for the Manhattan effect \cite{Fleury_2016}. This would give us an estimate of the length of the string in the network, or ``universe'', rest frame. We can also use those points to visualise the string. 
A disadvantage of this method is that it is possible to construct plaquettes with winding but no energy density with 
the Wilson energy functional used in the discretisation \cite{Hindmarsh:2017qff}. 

Another possible way of detecting strings is by searching for lattice points with a low value of the modulus of the field $\phi$. In most of the volume the field will be in its vacuum $|\phi|=\phi_0$, except for close to the string core. We use this estimator also to visualise the strings in the simulation, for which we set $|\phi_{\textrm{th}}| = 0.2\phi_0$ as the threshold. 

The main length estimator that we use during this work estimates the rest-frame length of the string, which takes into account all its energy components: potential, gradient and  kinetic. Strings obeying the Nambu-Goto equations would have a constant rest-frame length.

This length estimate relies on the fact that we can weight energies by functions that select regions of space occupied by string. Our choice of such function is the Lagrangian density. 
For example, we define the Lagrangian-weighted potential energy density as
\ben
E_{V,\mcL}=-\int {\rm d}^3x  V(\phi) \Ltilde\,.
\een
The subscript  $\mcL$ denotes that the quantity has been weighted by the dimensionless Lagrangian
\ben
\Ltilde = L/\phi_0^4,
\een
where $L$ is the Lagrangian (\ref{lagrangian}).

Following \cite{Hindmarsh:2017qff}, we use Lagrangian-weighted quantities to arrive at a length estimator  
\bea
\Llag &=&  \frac{1}{\mu_\mcL}\frac{E_\mcL - \Df L_\mcL}{1 + \Df} \label{e:ell_lag}\,.
\eea
Here
\ben
\mu_\mcL = - \int dx_\text{s}dy_\text{s} \left( \half \bB^2_\text{s} + |\bD\phi_\text{s}|^2 + V(\phi_\text{s}) \right)\Ltilde\, ,
\een
is the Lagrangian-weighted mass per unit length of  a static string, 
\ben
\mu_\mcL \Df= - \int dx_\text{s}dy_\text{s} \left( \half \bB^2_\text{s} - V(\phi_\text{s}) \right)\Ltilde\, ,
\een
is the Lagrangian-weighted difference between the magnetic and potential contributions to the energy, and $E_\mcL$  and $L_\mcL$ are the Lagrangian weighted Energy and Lagrangian, respectively.  The subscript `$\text{s}$' denotes that the coordinates are written, and the fields are measured,  in the local rest frame of the string. The constant $\Df \simeq 0$ as $\dx\to 0$ (see \cite{Hindmarsh:2017qff} for details).  

One shortcoming of $\Llag$ is that it estimates the total length of string inside the simulated volume, and it is not easy to use it to calculate the length of individual loops in the simulation. There are ways of dealing with this, shown in the following section.

Finally, using a similar procedure, one can write down an estimator for the mean-square velocity \cite{Hindmarsh:2017qff}:
\ben
\vLag^2=\frac{E_\mcL + L_\mcL}{E_\mcL - \Df L_\mcL} \,,
\label{vest}
\een
which will be used to estimate the velocities of the loops in the different simulations.


\section{Simulation setup and procedure}

We perform two different types of simulations during this work. One of them is a string network simulation,
in which we study string loops formed during the evolution of a network. The other type of simulation  consists of artificially setting up a string configuration which leads to the formation of loops 
engineered to follow a particular simple trajectory, likely to be well-described by the Nambu-Goto equation. 

In both types of simulations, we want to estimate the lifetime of a loop  and their velocity. 
We record the time at which the loop is formed $\tIni$ by visual inspection of  the string points, and their initial length $\LlagIni$.
We also need the time at which the loop disappears or evaporates $\tEva$. If the loop under study is the last one disappearing, we set $\tEva$ to be the instant at which $\Llag=0$. 
If the main loop vanishes and there is still another one in the simulation box, we obtain $\tEva$  by visual inspection. 
The lifetime of a loop, $\tLife$,  is given by
 \begin{equation}
 \tLife = \tEva - \tIni\, .
 \label{tlife}
 \end{equation}
 It is important to note that loops coming from string network simulations can fragment during their lifetime. In these cases we consider the daughter loops as continuations of the parent loop, and therefore we choose $\tEva$ to be the time at which the last one of the daughter loops disappears.
 
 We now describe the procedures of simulating network loops and artificial loops to obtain their $\tLife$, $\LlagIni$ and $\vLag$. All quantities with units of length and time are given in units of $\phi_0^{-1}$.
 
\subsection{Network loops}

The cubic lattices have  $N$ points per side and lattice spacing $\dx$ (and therefore lattice size $L=N\dx$). In the initial field configurations of network simulations only the scalar field is non-zero, and it is set to be a stationary Gaussian random field with a power spectrum 
\ben
P_\phi(k) = A e^{-(k \cl)^2}
\label{IC}
\een
where $A$ is chosen so that $\vev{|\phi^2|} = \phi_0^2$, and $\cl$ the correlation length, which we vary in different simulations.

Early phases of the simulation contain a considerable excess energy induced by the random initial conditions. We therefore  smooth the field distribution by applying a  period of diffusive evolution according to Eqs.~(\ref{AHdiff}). The diffusion period
starts at the beginning of the simulation at $\tStart=50$ and  ends at the time $\tDiff=70$. 
The time step during this diffusive period depends on the resolution used and is given by  $\delta t=(2/15)\dx^2$. After the diffusion period, the network evolves following the true equations of motion (\ref{eom}) and the time step used is $\delta t=(1/5)\dx$. It is during this evolution that the search for loops, and the estimation of their lifetime and velocity, is carried out. 

As we aim to study the full lifetime of loops until their evaporation, in this work we let the simulations evolve farther than half light-crossing time, which is the standard cutoff time for network simulations with periodic boundary conditions. Typically we set the final time of the simulations as three light-crossing times.

String networks formed from randomly generated initial conditions do not always necessarily end in decaying loops, despite the lattice's periodic boundary conditions. The topology of such lattices is the one of a torus, hence strings can wrap it and topology will forbid their decay.

 \begin{figure}[h]
    \centering
    \includegraphics[width=\columnwidth]{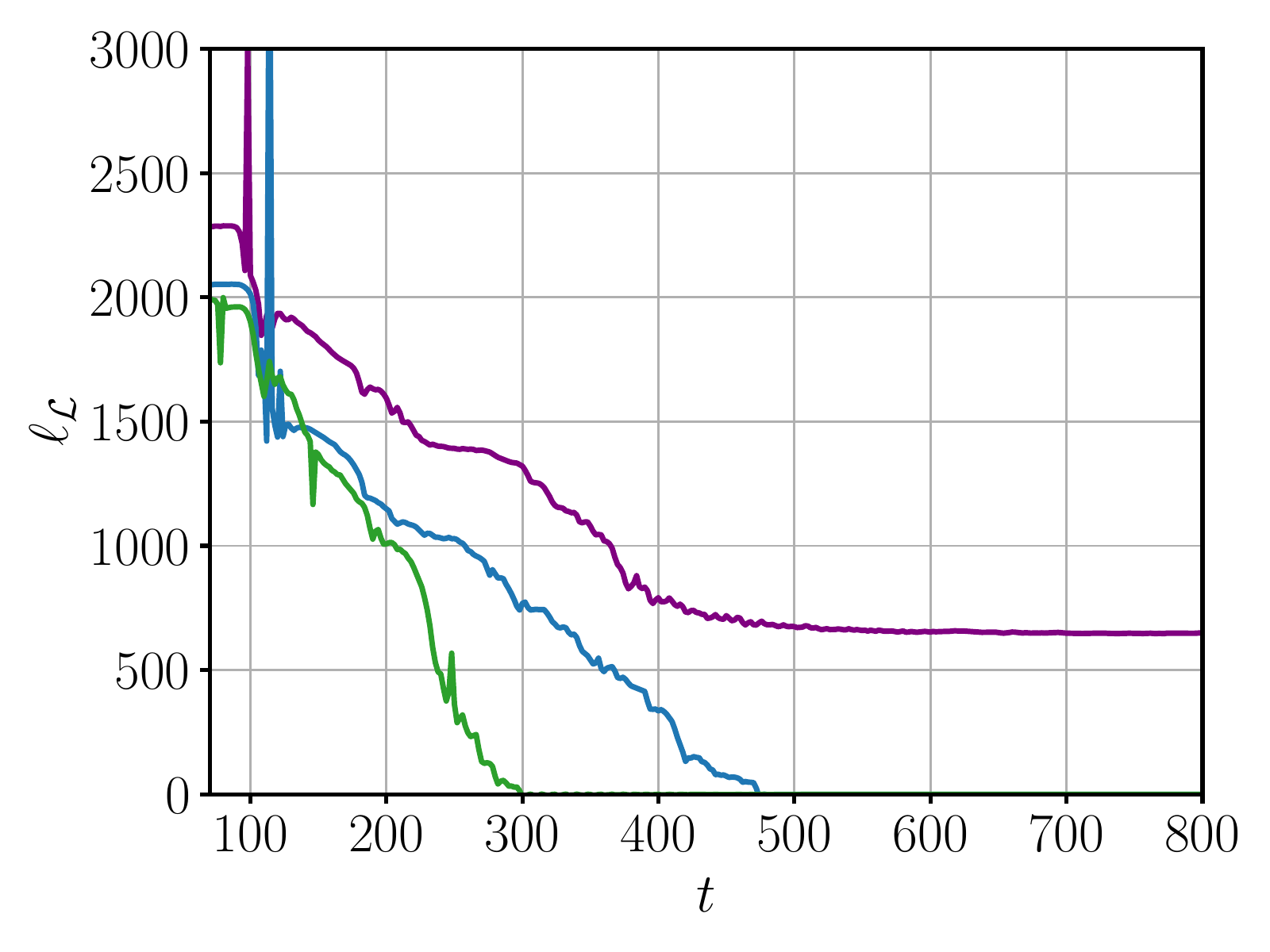}
    \caption{\label{fig:net_loopvsinfty}
    Comparison of the evolution of the invariant length estimator for simulations ending in a decaying loop formed by intersections (blue), in a decaying loop which is initially present and avoids intersections (green) and in strings that wrap around the periodic lattice (purple). All the cases correspond to   simulations with $N=1024$, $\dx=0.25$ and $\cl = 50$.  }
 \end{figure}

We focus on those runs where the simulation ends by loop decay. In order to find which initial conditions lead to that case, we follow the evolution of the invariant length estimator during the evolution. 

Fig.~\ref{fig:net_loopvsinfty} shows the evolution of the rest-frame string length for a  simulation with $\cl=50$ in a box of size $N=1024$ with $\dx=0.25$ (and hence $L = 256$). The green and blue lines correspond to simulations where there is no string left in the box by $t \approx 300$ and $t \approx 500$ respectively, indicating that the network has ended in a collapsing loop. 
The purple line shows a case in which the length decreases at approximately the same rate at the beginning, and once it reaches a scale of about $2L$, it stops losing energy. This case corresponds to that of two strings wrapping the torus\footnote{One cannot always conclude how many strings wrap the torus directly from the length estimator. Only a spatial visualisation of the network allows that.}.  Table~\ref{tab:summary} contains a summary of the number of simulations for each case. As it can be seen, the majority of simulations end in strings that wrap the box.
Of the total of 98 simulations with random initial conditions, 45 end in loops which completely evaporate.

 It is possible that a loop present in the initial state entirely avoids intersections, for example the one denoted by the green line in Fig~\ref{fig:net_loopvsinfty}.  These loops collapse at the same rate as the others, but are presumably unrepresentative of a scaling network, since they were created by the initial conditions. 
 These are more likely to form when the initial correlation length is large compared with the lattice size, and therefore, we choose $\cl$ to avoid this case.  Nonetheless, 14 of the 45 evaporating loops did not undergo an intersection, 
 and were not included in the analysis. 

\begin{figure}[h]
    \centering
    \includegraphics[width=\columnwidth]{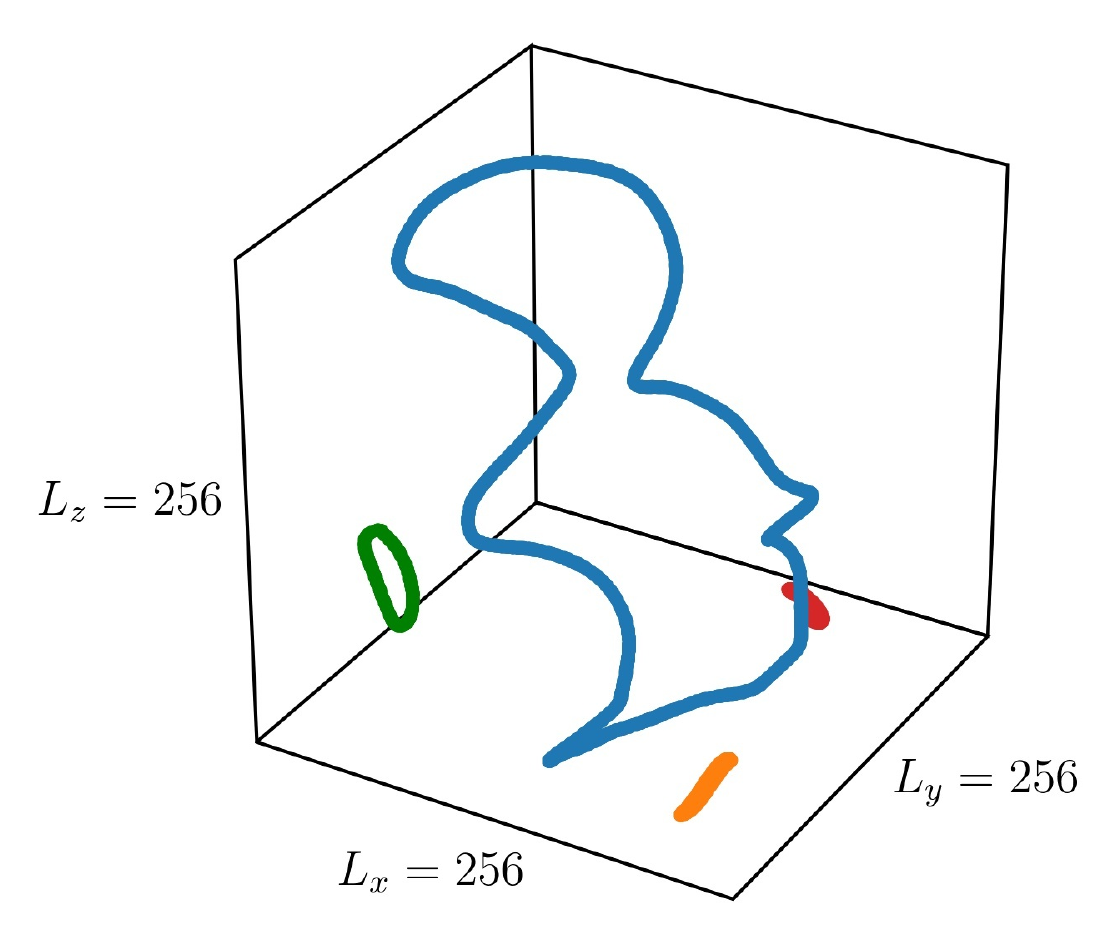}
    \caption{\label{net}
    Typical simulation of loops formed during network evolution, where there is one big main loop, and some smaller loops which evaporate quite early in the simulation.
    }
 \end{figure}

We have mentioned that the $\Llag$ estimator estimates the total length in the box, not only of the loop of interest. That is not a significant problem, because most of the length is in the main loop. 
In Fig.~\ref{net} we show a typical situation in these simulations, formed by a main big loop and some smaller loops that typically evaporate rather fast. The peaks that can be observed in the length estimate in Fig.~\ref{fig:net_loopvsinfty} correspond to  those small loops disappearing.

\begin{figure}[h]
    \centering
    \includegraphics[width=\columnwidth]{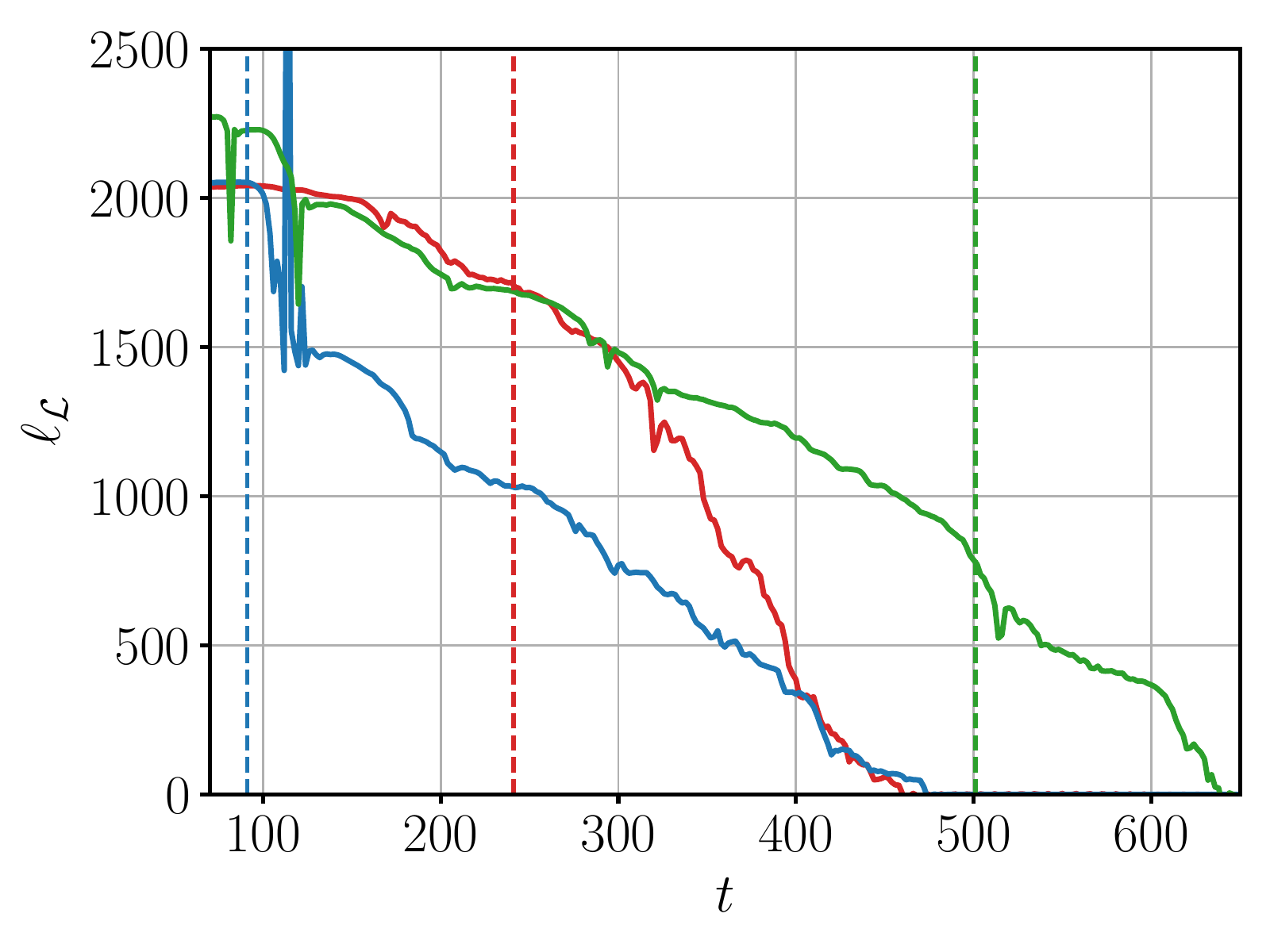}
    \caption{\label{fig:net_ellvst}
    Evolution of the Lagrangian weighted length estimator for three simulations with $N=1024$, $\dx=0.25$ and $\cl = 50$. Different colours stand for individual realisations and the  time of formation of the loop is represented with dashed vertical lines.}
 \end{figure}

Fig.~\ref{fig:net_ellvst} shows the rest-frame length of three  simulations, with $N=1024$, $\dx=0.25$ and $\cl=50$, ending in collapsing loops. The vertical dashed lines correspond to the time ($\tIni$) at which the final loop is formed for each case. 
 
 Note that  only a posteriori it is  known whether a given simulation will end up in a series of infinite strings or in a collapsing loop. Once the simulation is run, and it has been found that it does not end up in an infinite loop, we visualise the strings using  the scalar phase winding in different plaquettes to determine the time of formation of a loop by visual inspection.

\begin{figure}[h]
    \centering
    \includegraphics[width=\columnwidth]{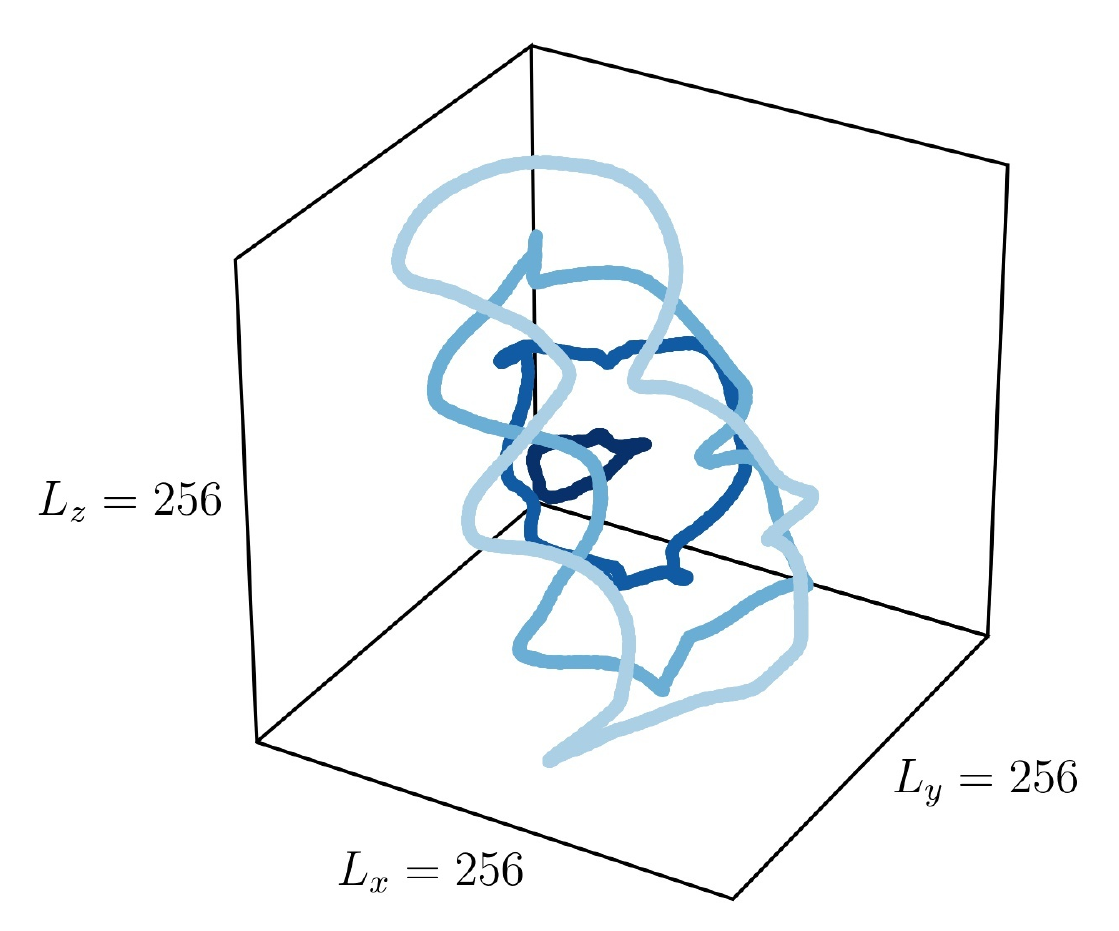}
    \caption{\label{fig:snap_net}
    Sequence of snapshots of the evolution of the loop represented by the blue line in Figs.~\ref{fig:net_loopvsinfty} and \ref{fig:net_ellvst}. Lighter blue corresponds to earlier times and darker to later times. The first time step corresponds to  $t=92$ and the steps are taken at equally time steps with $\Delta t = 100$.}
 \end{figure}

 A typical evolution of a loop formed by the intersection of strings in the network is shown in Fig.~\ref{fig:snap_net}, which corresponds to the loop in the blue line in Figs.~\ref{fig:net_loopvsinfty} and \ref{fig:net_ellvst}. We plot 4 different snapshot of the points with $|\phi| <  0.2 \phi_0$, 
taken at equally time-spaced moments ($\Delta t =100$) between  $t=92$ and $t=392$,  from lighter to darker blue. 
The video corresponding to the full evolution of this loop can be found in \cite{AHLoopVideos}.
 
\subsection{Artificial string configurations}
\label{subsec:AI}

We also study artificially created string configurations.  We aim to obtain loops that live long enough to oscillate in a Nambu-Goto-like trajectory, as demonstrated to exist in \cite{Matsunami:2019fss}, and compare them with network-produced loops. 

Our method for constructing oscillating loops is slightly different from Ref.~\cite{Matsunami:2019fss}, as we 
form strings which are initially stationary, and allow their self-intersection to generate the loop.  
This avoids the need to boost the strings after the diffusive phase.

The initial string curves are specified as piecewise smooth functions, and mapped to cells of the lattice.
Adjacent cells are separated by plaquettes $\{ {\cal P} \}$, whose links are all 
set to $\pi/2$, thus generating a flux of $2\pi$ through the plaquette.  
All other links are set to zero, and the scalar field $\phi$ is set to the vacuum expectation value everywhere. The configuration is then cooled down by evolving the system under a diffusive phase (\ref{AHdiff}) for $5$ time units. 
The time step used is the same as one the for network simulations.

The two ingredients we use to set up the configuration of these artificial initial conditions are sinusoidal and sawtooth standing waves wrapped around the simulation box. 
The sinusoidal standing wave lying on the $(x,z)$ plane, with amplitude $A$, 
had coordinates $(X,Z)$ given by  
\ben
X = A  \cos( 2\pi Z/L) .
\label{eq:SWStrCur}
\een
The sawtooth string lying in the $(x,y)$ plane with amplitude $A$ had coordinates 
\ben
X = \begin{cases} 
     \displaystyle A\Big[\frac{Y}{(L/4)}-1\Big], & 0\leq Y\leq L/2 \\[2ex]
     \displaystyle  -A\Big[\frac{Y}{(L/4)}-3\Big], & L/2 < Y < L
   \end{cases} \, .
\label{eq:KiStrCur}
\een

\begin{figure}[h]
    \centering
    \includegraphics[width=\columnwidth]{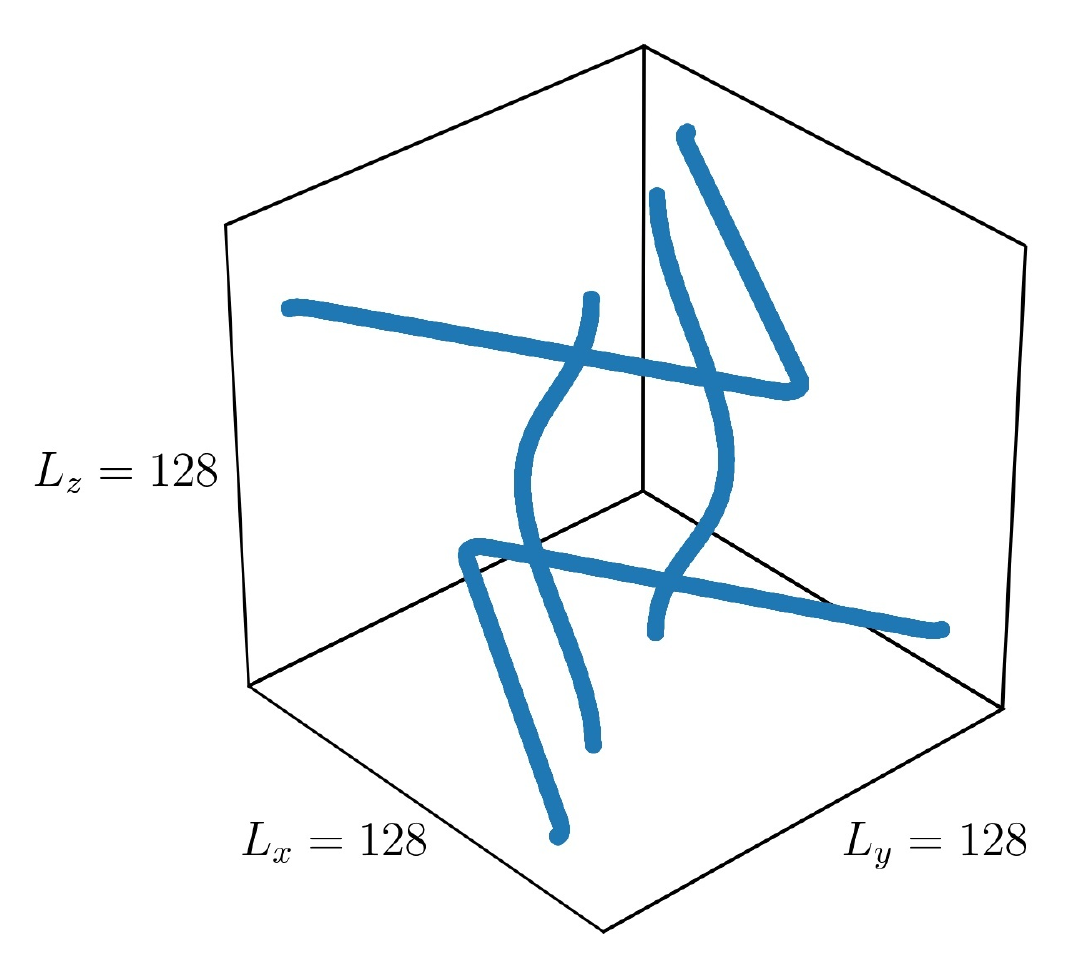}
    \caption{\label{fig:IC_art_loop}
    Initial configuration to create artificial loops: two sawtooth strings with kinks oriented in opposite directions with $A=0.5L$ and two strings with sinusoidal standing waves with $A=0.1L$. 
}
 \end{figure}

We first experimented with configurations consisting of two straight lines interacting with two sawtooth strings, but we observed that in those cases after recombination we ended up having a double antiparallel straight line, 
which immediately annihilated.

In order to get oscillating, or longer lasting loops, instead of using kinks we started combining segments of strings with sinusoidal waves on them. These also led to double lines, and if we tried to increase the amplitude of the standing wave, the recombinations happened much earlier than desired, and there was no long-lasting loop left.

Finally, we obtained the desired object by carefully combining both types of standing wave.  One example of such configurations is shown in Fig.~\ref{fig:IC_art_loop}. In this example, both sawtooth strings have amplitudes $A=0.5L$ and lie in the $(x,y)$ plane at different values, $Z=L/10$ and $Z=9L/10$ respectively. The kinks are oriented in opposite directions and thus they will move in opposite directions. The two sinusoidal strings, in turn, lie the $(x,z)$ plane at $Y=L/3$ and $Y=2/3L$, are set in anti-phase and have a lower amplitude $A=0.1L$. The magnetic flux that runs through the strings is set up in such a way that when intersection happens, the desired loop is created. In order to achieve this, the flux flows in opposite directions through the pairs of strings. In the case of the sawtooth strings, for the upper string in Fig.~\ref{fig:IC_art_loop} the magnetic flux runs in the $y$ direction, while in the lower string in the $-y$ direction. Similarly, in the standing wave on the left-hand side the direction is $z$ and for the one on the right-hand side $-z$.

This initial configuration leads the corners on the sawtooth to resolve into two kinks travelling with the speed of light in 
opposite directions, separated by straight segments of string oriented parallel to the $y$ axis. 
The sinusoidal waves oscillate approximately as a classical standing wave on a non-relativistic stretched string. 
At some point,  
the straight segments on the sawtooth strings collide with the sinusoidal standing waves and two loops are created from the intersection: one inner loop and an outer one (due to the periodic boundary conditions). We have found that loops last longer if the initial conditions are chosen such that  the collisions do not happen at the same moment.  This manner of preparing loops prevents the loop from disappearing due to the double-line annihilation, and allows them to oscillate.

In Fig.~\ref{fig:IC_art_loop} it can be seen that the distance between strings  is maximised so that the size of the inner loop is as large as possible. Nonetheless, in the case of the sinusoidal waves the distance between them needs to be carefully chosen so as to avoid possible early intersections with the sawtooth strings. For the same reason, the amplitude of the sinusoidal waves cannot be as high as for the  sawtooth strings.

\begin{figure}[h]
    \centering
    \includegraphics[width=\columnwidth]{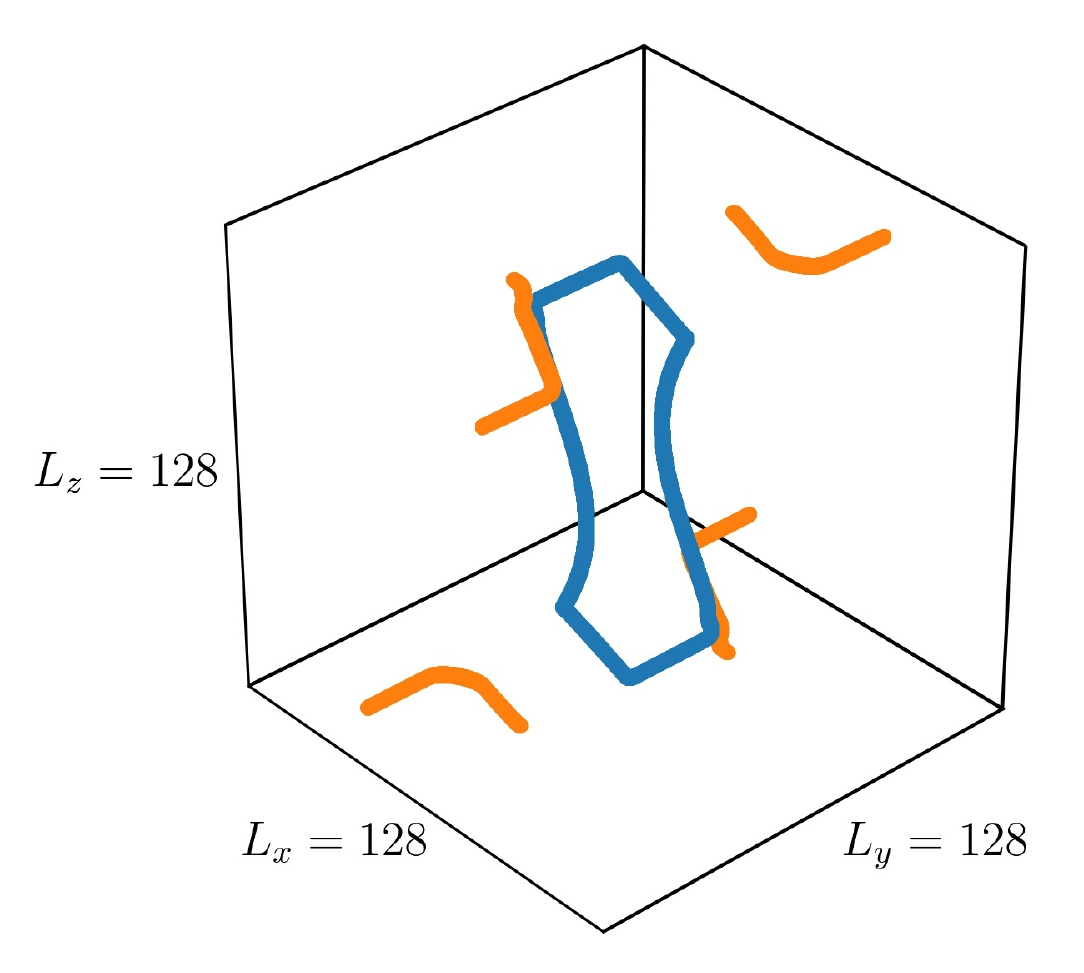}
    \caption{\label{2loops}
Simulation of the artificially created loops: typically there are two loops created, of roughly the same length.}
 \end{figure}

The length of the string is estimated by the Lagrangian weighted length estimator. As before, the length that the estimator gets is not only  the length of the loop under study, but all the length in the simulation. In the network case we saw that most of the length was in the main loop. That is not the case here, since the two loops formed are roughly of the same length. Therefore we can approximate the length of the loop as roughly half the estimate given by $\Llag$. In Fig.~\ref{2loops} we can see the typical situation at formation of the artificial loops.

\begin{figure}[h]
    \centering
    \includegraphics[width=\columnwidth]{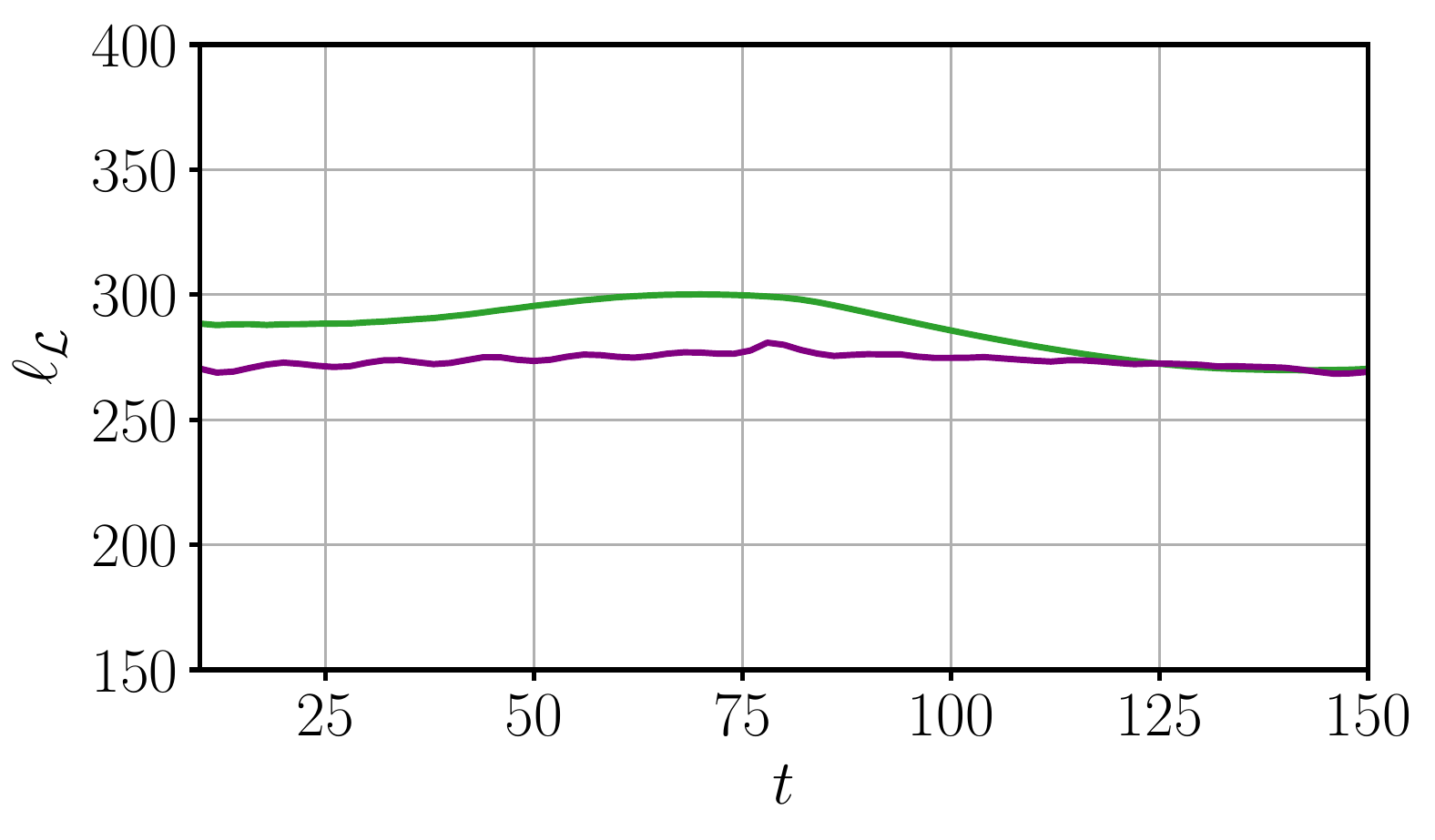}
    \caption{\label{fig:SWKinky_length}
    Evolution of the length estimator for standing waves starting in a sawtooth curve (purple) and sinusoidal curve (green), both with  $A=0.5L$ (see Eqs.~\ref{eq:SWStrCur}, \ref{eq:KiStrCur}), over one period.}
 \end{figure}

Before studying the evolution of these loops, we perform a preliminary test on the length estimator for lattices with individual standing waves and sawtooth strings with  amplitudes  $A=0.5L$. Fig.~\ref{fig:SWKinky_length} shows the evolution over a period of oscillation of the invariant length estimator for a simulation of a single oscillating standing wave for a  simulation with $N=1024$, $\delta x = 0.125$ (green line). The purple line corresponds to a similar simulation of a single sawtooth string. This figure shows that the invariant length is well conserved for both cases, though small variations (below $10\%$) are present for the standing wave. 

 \begin{figure}[h]
    \centering
    \includegraphics[width=\columnwidth]{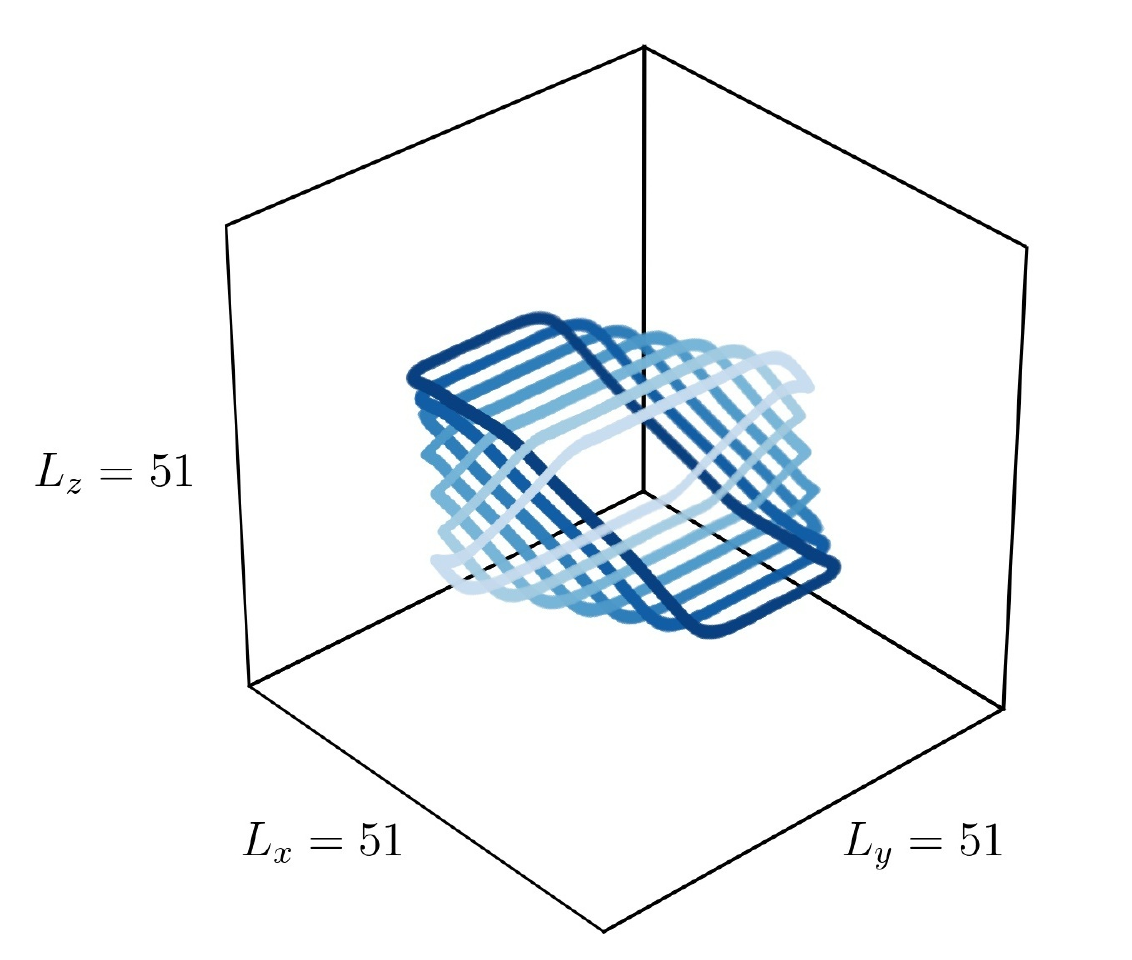}
    \caption{\label{fig:arti_sequence}
    Time sequence of half an oscillation of an artificially prepared loop shown for a box with length $L=128$, taken from a $N=1024$ simulation with $\delta x = 0.125$. Line colours go from darker to lighter, and the time step is $\Delta t = 4$.}
 \end{figure}

Fig.~\ref{fig:arti_sequence} includes several snapshots of the typical evolution of an artificially prepared loop. In this case, the loop was  simulated in a lattice with $N=1024$, and $\delta x = 0.125$ (therefore, a box of sides $L=128$) . We plot half of an oscillation period, going from lighter blue to darker blue, also including intermediate steps separated by $\Delta t = 4$. These loops evade successfully the typical double-line collapse and start rotating and oscillating. 
The video corresponding to the formation and full evolution of this loop can be found in \cite{AHLoopVideos}.

\section{results}

In this section we present the results of the two sets of simulations performed, both for loops coming from network evolution and from artificially created loops.

\begin{table}[h]
\begin{tabular}{|c|c|c|c|c|c|c|}
\hline
\multicolumn{1}{|c|}{$\cl$} & \multicolumn{1}{c|}{$N$} & $\dx$ & Runs & $\infty$ strings &   from ICs & from intersection \\ \hline
100 & 2048 & 0.25 & 6 & 2 & 4 & 0 \\ \hline
50 & 2048 & 0.25 & 5 & 3 & 0 & 2 \\ \hline
50 & 1024 & 0.25 & 15 & 9 & 3 & 3 \\ \hline
25 & 1024 & 0.25 & 22 & 14 & 2 & 6 \\ \hline
15 & 1024 & 0.25 & 10 & 5 & 0 & 5 \\ \hline
10 & 1024 & 0.125 & 10 & 3 & 0 & 7 \\ \hline
50 & 1024 & 0.125 & 4 & 2 & 2 & 0 \\ \hline
25 & 1024 & 0.125 & 4 & 0 & 2 & 2 \\ \hline
15 & 1024 & 0.125 & 22 & 15 & 1 & 6 \\ \hline
\multicolumn{1}{|c}{}  & \multicolumn{1}{c}{Total}  & \multicolumn{1}{c|}{} & \multicolumn{1}{c|}{98} & \multicolumn{1}{c|}{53} & \multicolumn{1}{c|}{14} & \multicolumn{1}{c|}{31} \\ \hline
\end{tabular}
\caption{Summary of the different runs used to study the  evolution of loops created in networks. The simulations have been performed in cubic lattices with $N$ points per dimension and lattice-spacing $\dx$. The initial correlation length was given by $\cl$ (see Eq.~\ref{IC}). We have performed a number of runs (Runs) of each type.  Of those runs,  those that ended up as infinite strings ($\infty$ strings) or where loops were already present in the initial conditions (ICs) were not included in the lifetime analysis.
\label{tab:summary}}
\end{table}

The network simulations were performed in periodic lattices of different number of points $N$ and lattice-spacings $\dx$, which we summarise in Table~\ref{tab:summary}. The initial configurations (\ref{IC}) have a variety of initial correlation lengths $\cl$. We performed a number of runs of each type (varying the initial random configuration) and studied the fate of the loops in each case. Resolution tests can be found in the Appendix.

Just over half of the simulations ended in infinite strings. 
In some other simulations, the loops were already present in the initial conditions (ICs), and did not undergo 
an intersection before evaporating.  

\begin{figure}[h]
    \centering
    \includegraphics[width=\columnwidth]{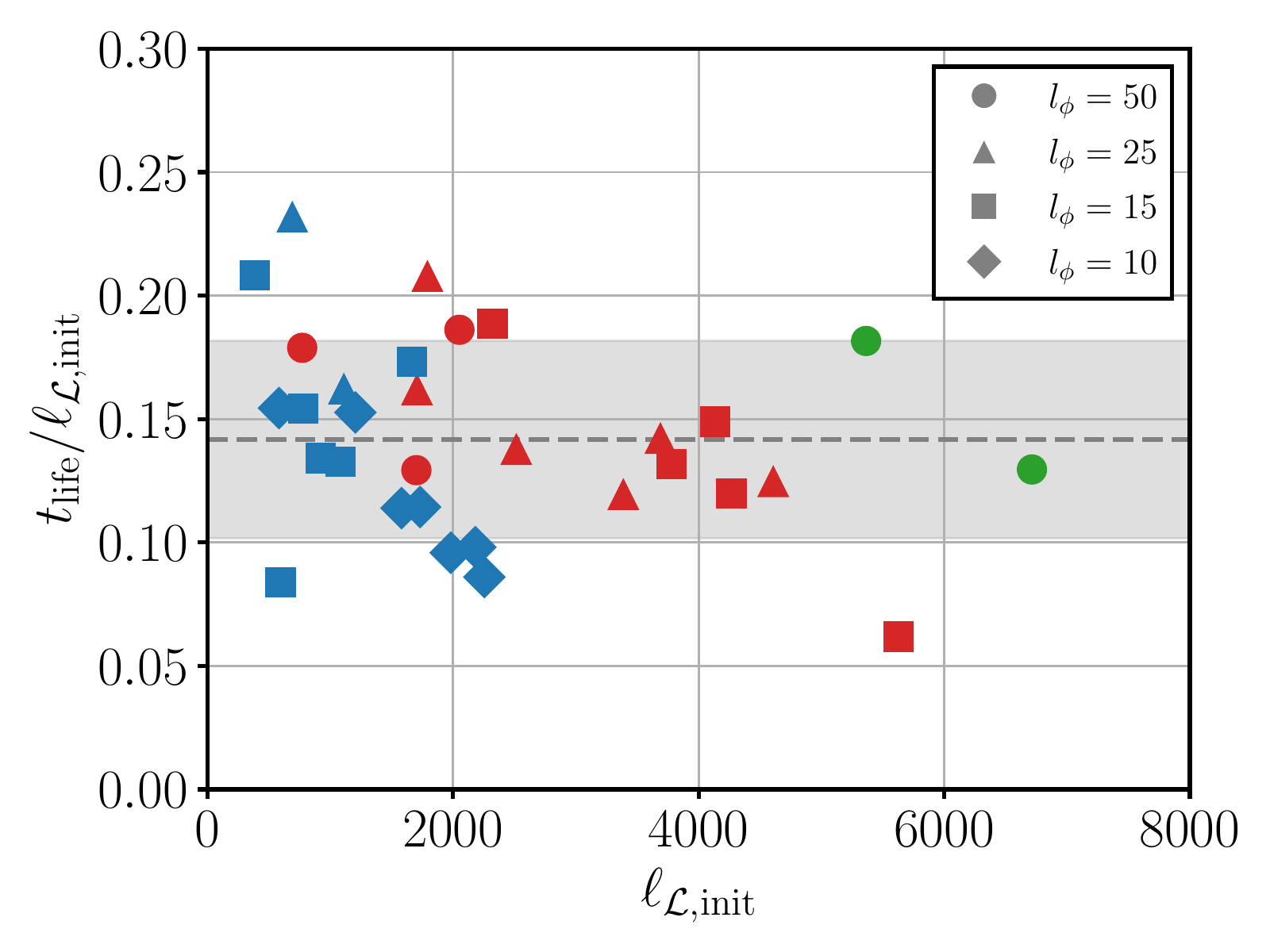}
    \caption{\label{fig:net_tauvsell}
    Scatter plot of the lifetimes of loops from network simulations for the whole set of lattice sizes, ($N=1024$, $\dx=0.125$) (blue), ($N=1024$, $\dx=0.25$) (red) and ($N=2028$, $\dx=0.25$) (green). Different markers correspond to different correlation lengths in the initial conditions. The horizontal dashed line represents the overall mean value of the lifetime, and the shaded region represents the 1-$\sigma$ band.}
 \end{figure}

In 31 of the 98 simulations large loops were created by intersections of the network, which were taken as 
the best model of loops created by a scaling network evolution. 
For these loops we recorded the initial length of the strings $\LlagIni$, the time they were formed and the time where they disappeared to estimate their lifetime $\tLife$ (\ref{tlife}). Fig.~\ref{fig:net_tauvsell} shows the relation between the size of the loops and their decay rate, represented as the lifetime divided by $\LlagIni$. We use blue markers for lattices with ($N=1024$, $\dx=0.125$), red for ($N=1024$, $\dx=0.25$) and green for ($N=2028$, $\dx=0.25$). Different shapes stand for different initial correlation lengths. 

First we observe that we obtain some rather large loops, of $\LlagIni$ of up to 6000 inverse mass units. 
Secondly, we observe that, as expected from previous works on field theory simulations \cite{Hindmarsh:2008dw,Hindmarsh:2017qff}, loops formed at randomly generated networks do not live for a long time in comparison to their initial length.  

Loops that are initially larger live longer, as expected, but when we normalize their lifetime with the initial length, we see no change in the trend of the decay rate. In fact, 
for all loops simulated in this work the points representing the decay rate tend to cluster around a constant value, regardless how large they are. Therefore, we find that for this kind of loops the decay rate scales approximately as $\propto \LlagIni$ for all initial sizes of loops.   We compute the proportionality constant by averaging over all cases, giving $0.14\pm0.04$, represented as a gray dashed line in the figure, together with the 1-$\sigma$ band.

This contrasts with the scaling reported in \cite{Matsunami:2019fss}, \ie $\propto \LlagIni^2$, presented for loops artificially created by colliding four straight strings wrapping the simulation volume.  In order to understand this difference, we perform a series of simulations with artificially created initial conditions (as explained above), to show that our model and numerical  setup allow for long lasting loops. 
We performed a series of runs to check for parameters and resolution tests, which can be found in the Appendix,  
resulting in the choice of  $\dx=0.125$.
We have performed 8 production runs, where loops were created with a pair of sawtooth strings intersecting with a  pair of sinusoidal strings with amplitude $A$ (see subsection~\ref{subsec:AI}). Four of the simulations had initial  standing wave amplitude $A=0.1L$ and four initial amplitude $A=0.075L$. The lattice sizes used were $N=768, 1024, 1280, 1536$.

\begin{figure}[h]
    \centering
    \includegraphics[width=\columnwidth]{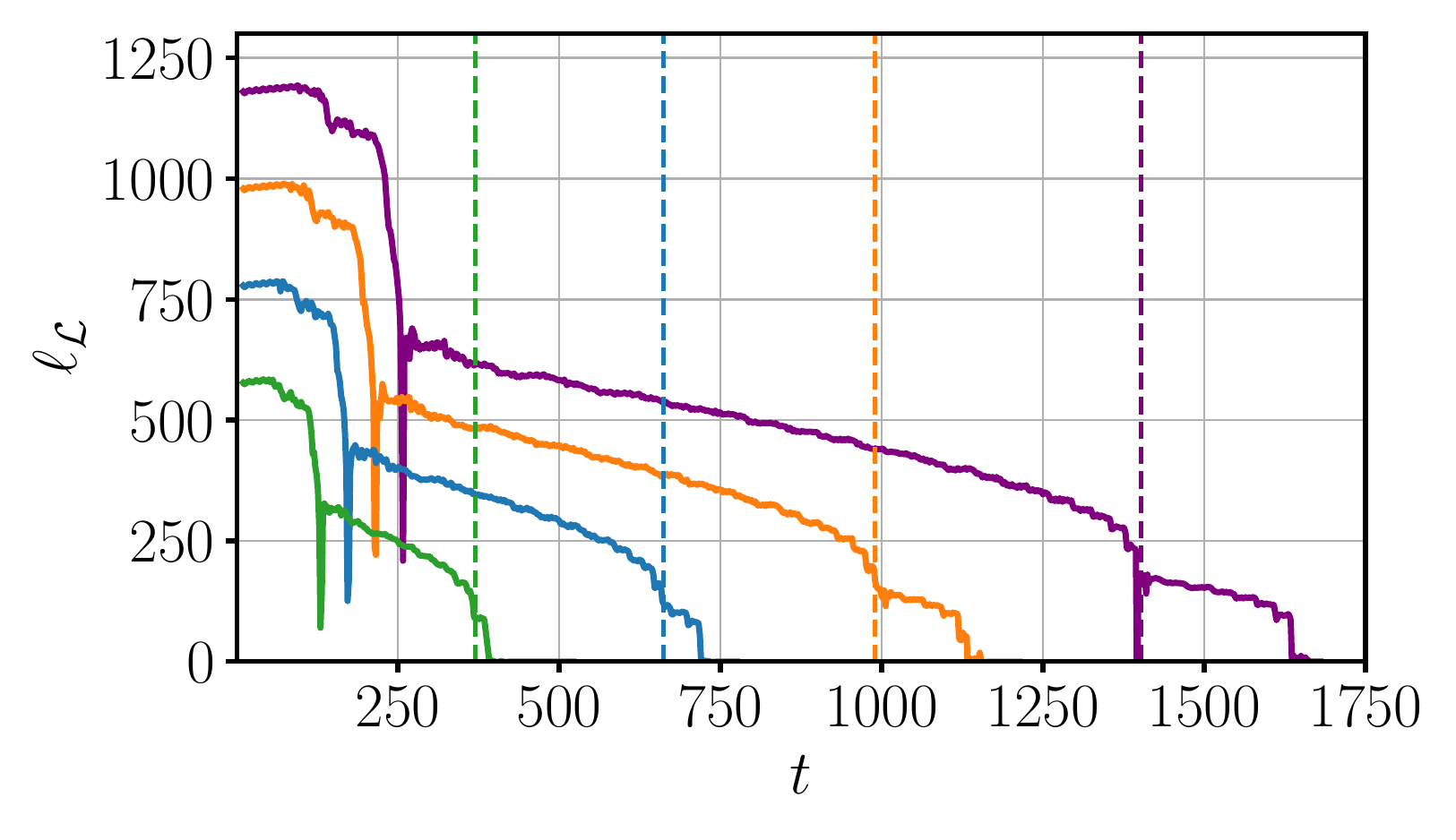}
    \includegraphics[width=\columnwidth]{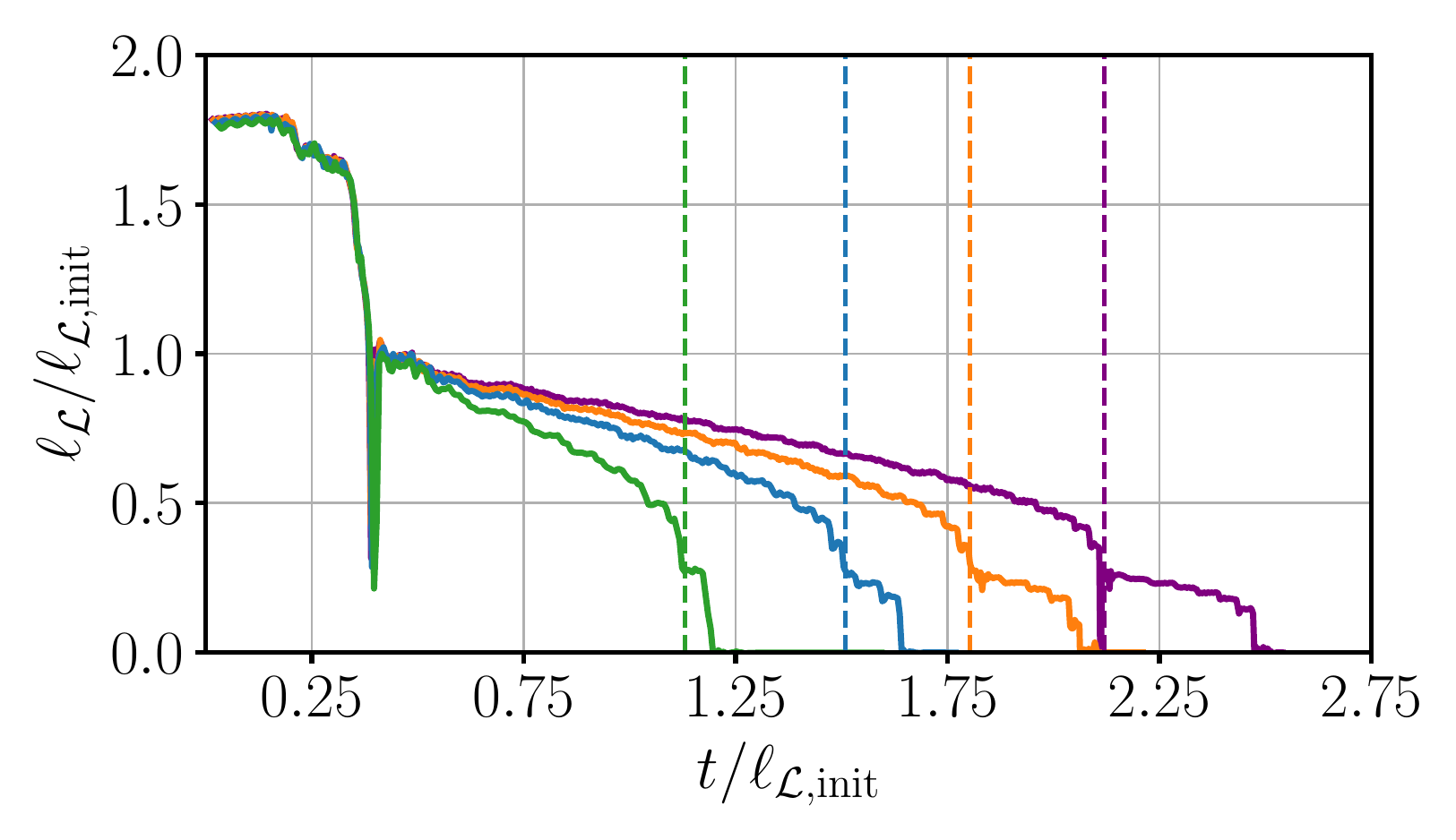}
    \includegraphics[width=\columnwidth]{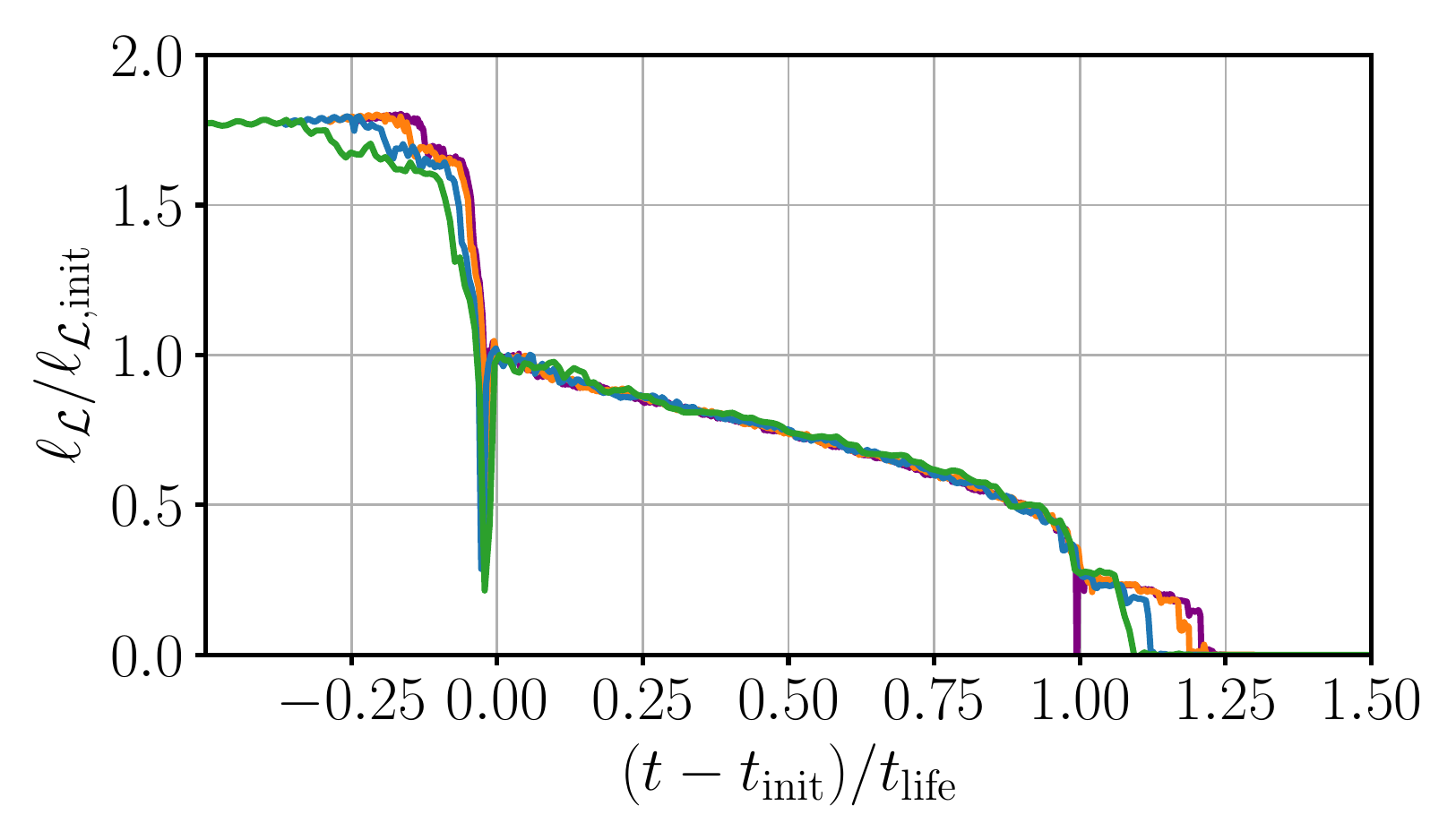}
    \caption{\label{fig:arti_normellvst}
    Evolution of the loop length estimator corresponding to artificial loops created from standing waves with amplitude equal to $A=0.1L$. Each colour represent different sizes for the lattices, green for $N=768$, blue for $N=1024$, orange for $N=1280$ and purple for $N=1536$. Dashed vertical lines represent the moment where the inner loop disappears. 
The top panel shows the un-normalized evolution of the length estimator. The middle panel shows the 
length estimator normalised to its value $\LlagIni$ at the beginning of the phase of slow decay, and the bottom panel shows the normalised evolution but with the time variable shifted and scaled 
so that the loop formation and disappearance coincide. 
 }
  \end{figure}

We show the evolution of the length estimator for the simulations with standing wave amplitude $A=0.1L$  in the upper panel of Fig.~\ref{fig:arti_normellvst}. The colour varies with the size of the lattice: $N=768$ (green), $N=1024$ (blue), $N=1280$ (orange) and $N=1536$ (purple).  As the figure shows, all cases have a similar behaviour.

At the beginning the length estimator remains constant for all cases, while 
the standing waves self-intersect and the loop starts to oscillate.  Then the loops are formed after which they lose approximately half their energy, independent of their size. 
Afterwards, the length decays much more slowly:
loops of different initial length show a different decay rate, the larger the loop the smaller the decay rate.

We consider the formation time of the loops $\tIni$ to be just after the end of the burst of energy loss 
following intersection, and 
determine the initial length of the loops ($\LlagIni$) as the length at that point.  
The disappearance time of the loops is estimated by visual inspection, shown by the dashed vertical lines in Fig.~\ref{fig:arti_normellvst}. The difference between the two is the lifetime $\tLife$.

The similarities of all cases are more clearly shown in the middle panel of  Fig.~\ref{fig:arti_normellvst}, where we plot the invariant length normalised to $\LlagIni$. All cases follow similarly shaped curves, from the beginning, through the formation of the loops, the loss of half of their energy, and the subsequent smooth evolution, from $t/\LlagIni\approx$  $0.41 \ (N=768)$, $0.42 \ (N=1024)$, $0.42 \ (N=1280)$, $0.42 \ (N=1536)$. This figure shows clearly the slowing of the decay with increasing initial length.

The bottom panel shows the same simulations, but with the time normalised in such a way that the 
curves are aligned at loop formation and the disappearance of the inner l
oop at $t_\text{life}$, i.e., we plot the normalized length versus $(t-
t_\text{init})/t_\text{life}$.  The collapse onto a single curve is remarkable.

\begin{figure}[h]
    \centering
    \includegraphics[width=\columnwidth]{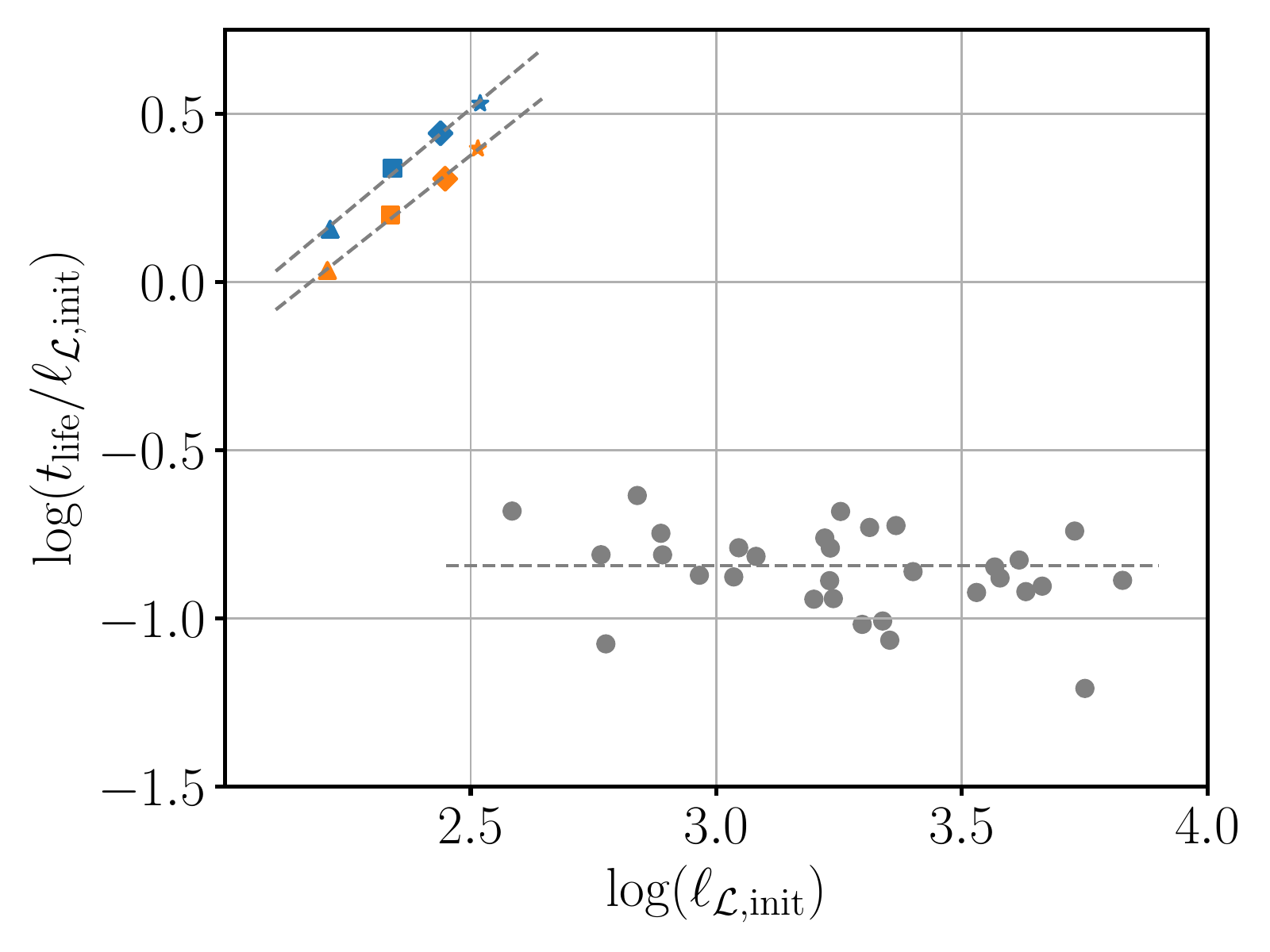}
    \caption{\label{fig:arti_powerlaw}
    Loop lifetime versus initial loop length for loops constructed from collisions between kinky strings and standing waves with amplitudes $A=0.1L$ (blue) and $A=0.075L$ (orange). Each marker represents different sizes for the lattices, triangle for $N=768$, square for $N=1024$, diamond for $N=1280$ and star for $N=1536$. Dashed lines are the fits of the data points by expression given in Eq.~(\ref{eq:arti_fit}). The points corresponding to loops created from intersection of infinite strings are also included (gray).}
 \end{figure}

The dependence of the loop lifetime on its initial length is seen in Fig.~\ref{fig:arti_powerlaw}, where we 
plot the normalised lifetime against the initial length of the loops.  
The left upper corner of Fig.~\ref{fig:arti_powerlaw} contains the points representing artificial loops. Blue markers correspond to loops prepared from standing waves with $A=0.1L$ and orange markers for $A=0.075L$. For comparison with Fig.~\ref{fig:net_tauvsell} we also include the collection of loops obtained from network simulations in gray.  

In order to determine how the lifetime depends on the initial length, we fit the four points for each amplitude using the following function:
\begin{equation}
\tLife\phi_0 = \alpha (\LlagIni \phi_0)^\beta\, ,	
\label{eq:arti_fit}
\end{equation}
where we have restored the dimensionful quantity $\phi_0$ for clarity. 
We find $\alpha = (3.0 \pm 0.4)\times10^{-3}$ and $\beta =2.22 \pm 0.06$ (for $A=0.1L$) and $\alpha = (3.0 \pm 0.3)\times10^{-3}$ and $\beta =2.16 \pm 0.05$ (for $A=0.075L$).  The fit is shown as a dashed line Fig.~\ref{fig:arti_powerlaw}.

The lifetime scaling power law index $\be$ is close to the value reported in \cite{Matsunami:2019fss} for loops consisting of straight segments and kinks.  The coefficient $\alpha$ is similar in magnitude. 
The construction and simulation of long-lasting loops is a therefore good consistency check. 
The decrease in energy of our loops is less episodic than those of Ref.~ \cite{Matsunami:2019fss}, 
indicating that energy loss is not just occurring when kinks collide.
 
We also studied the velocity of the loops, both formed in the network, and those created artificially, using the velocity estimator described in Eq.~(\ref{vest}).

\begin{figure}[h]
    \centering
    \includegraphics[width=\columnwidth]{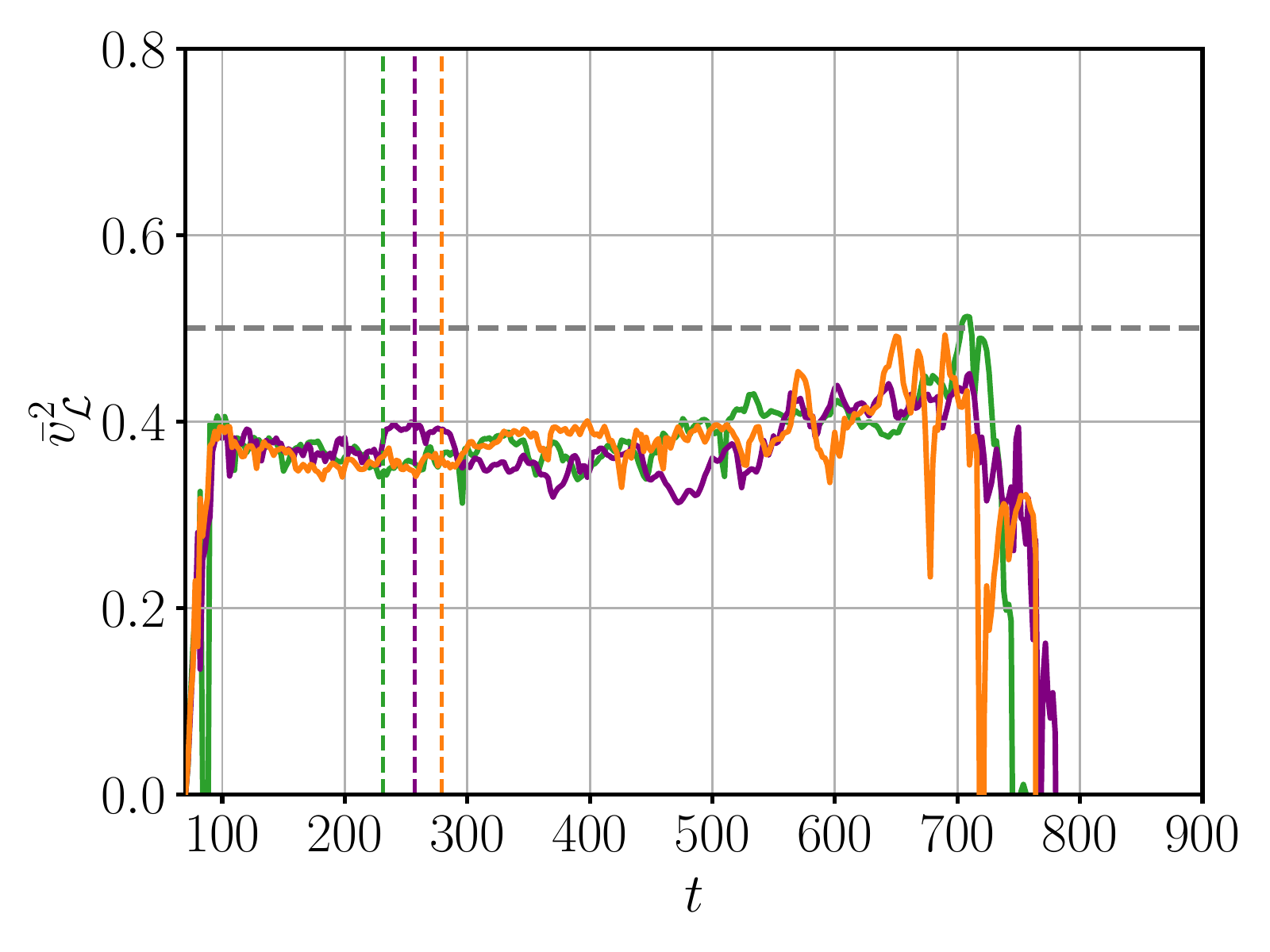}
    \caption{\label{networkvel}
    Velocity estimator plot for loops formed from network evolution. The colours correspond to three different runs with $\cl=15$, $N=1024$ and $\dx=0.25$.  The vertical lines show the moment of formation of the loop. The dashed grey horizontal line is the expected time-averaged mean square velocity of a Nambu-Goto loop in Minkowski spacetime.}
 \end{figure}

For loops formed during the network evolution, Fig.~\ref{networkvel} shows the typical behaviour of the mean square velocity $\vAv^2$. The three colours correspond to  three different runs, all with $\cl=15$, $N=1024$ and $\dx=0.25$.  The vertical lines show when the loop was formed.  The horizontal grey dashed line shows the expectation from Nambu-Goto dynamics, which is that $\vAv^2 = 0.5$ averaged over one oscillation, in Minkowski spacetime. 
It is very clear that the network loops' average velocity is rather lower in all cases.

\begin{figure}[h]
    \centering
    \includegraphics[width=\columnwidth]{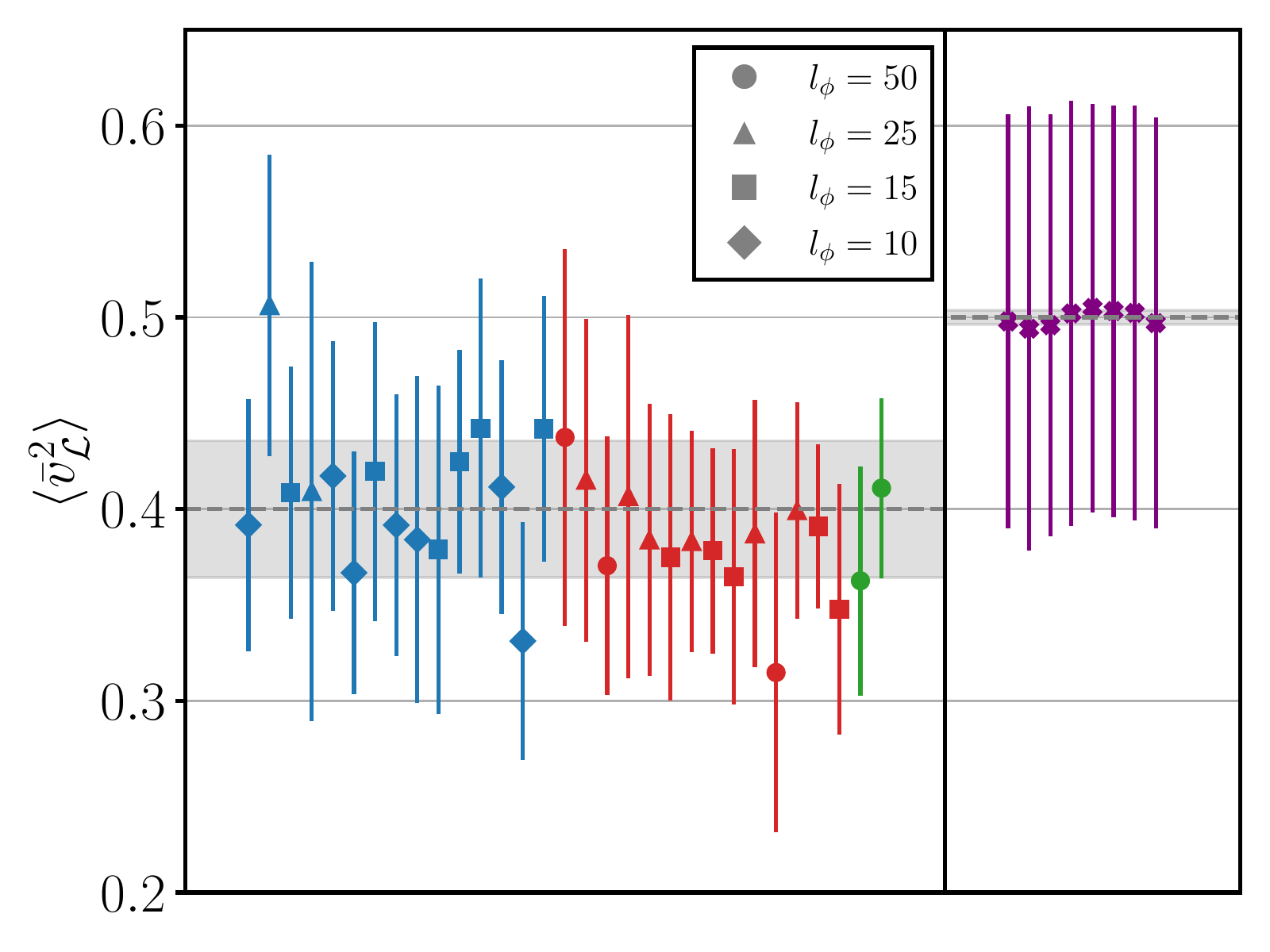}
    \caption{\label{allvel}
Time-averaged mean square velocities for all loops created by network dynamics (left panel) and by the artificial procedure (right panel). 
The error bars show the standard deviation over the lifetime of the loop. The dashed line correspond to the average of all the velocity values, and the grey area corresponds to the standard deviation of the mean. 
}
 \end{figure}

To make a more quantitative statement, 
we averaged the mean square velocity of each loop from its formation until its disappearance. The result can be seen in Fig.~\ref{allvel}, where we have plotted the time-averaged mean square velocity of each of the 31 loops created by network dynamics. The errors shown are the standard deviation from the time-average. The dashed line correspond to the average of all the velocity values, and the grey area corresponds to the standard deviation of the mean: $0.40\pm0.04$.   There is clearly a wide spread of the values of velocities, both from the calculation of the average velocity of a single loop, and also from simulation to simulation.  All (but one) of the loops are far from saturating the value predicted by Nambu-Goto dynamics.

\begin{figure}[h]
    \centering
    \includegraphics[width=\columnwidth]{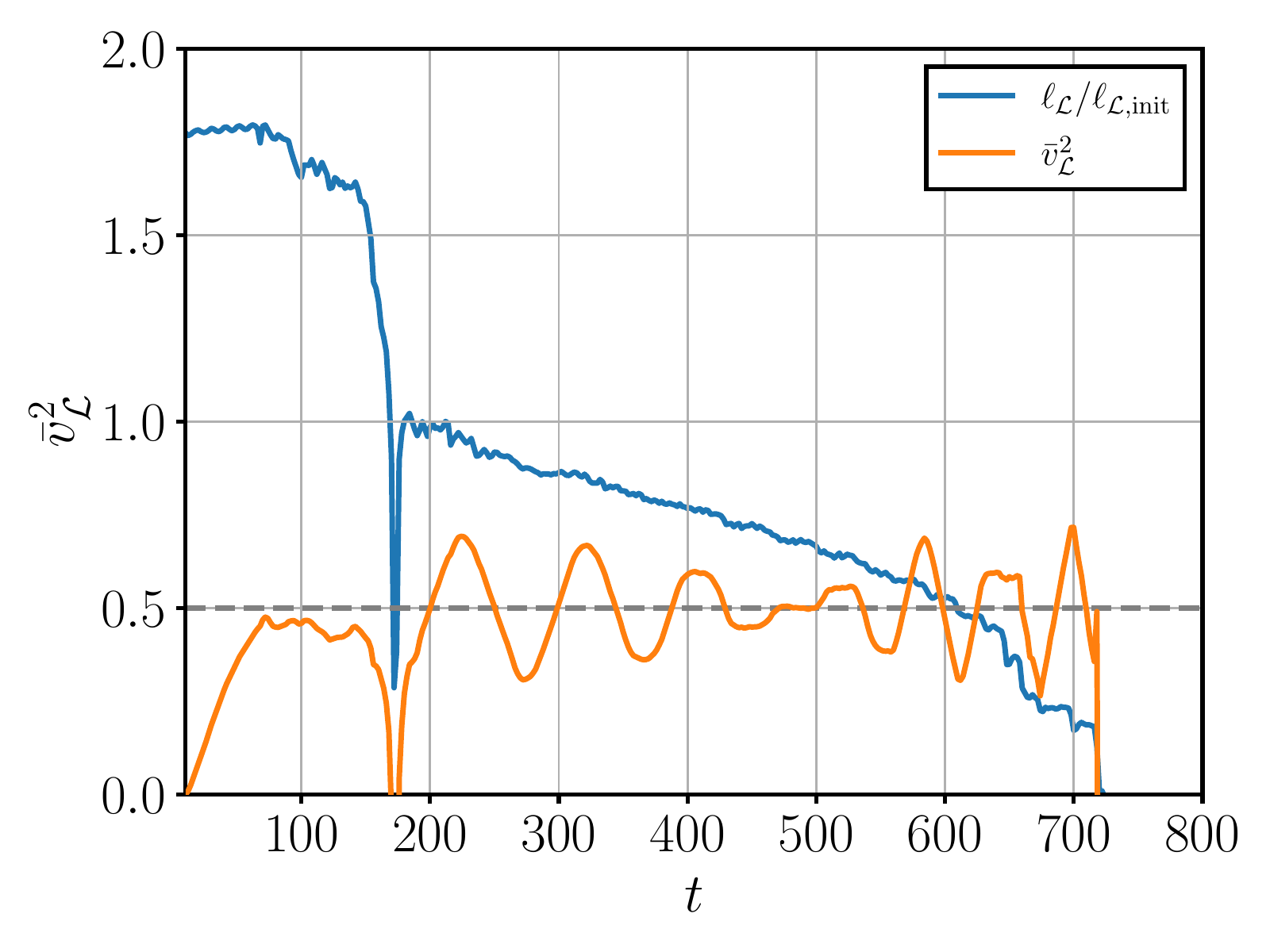}
    \caption{\label{artifvel}
    Length and mean square velocity for an artificial loop formed from the collision of a pair of sawtooth standing waves with amplitude $A=0.5L$ and a pair of sinusoidal standing waves, with $A = 0.1L$, as described in subsection \ref{subsec:AI}. The length is expressed as a fraction of its initial length. The dashed grey horizontal line is the expected time average of the mean square velocity for a loop following Nambu-Goto dynamics.}
 \end{figure}

This behaviour is in clear contrast with the artificial loops, as shown in Fig.~\ref{artifvel}. In the figure, we have shown the length and velocity estimators for a pair of artificial loops, formed with a wave of amplitude $A=0.1L$. The horizontal grey line also shows the Nambu-Goto behaviour in Minkowski spacetime, which in this case loops satisfy: the mean square velocity of the loops oscillates clearly around the Nambu-Goto prediction.  The fact that there are two loops accounts for the ``beating'' of the oscillation, with a 
node around $t = 450$. 
  
Fig.~\ref{allvel} also shows, in purple, the average values of the velocities for artificially created loops. The time-average is very close to the Nambu-Goto predicted value, with very little variation between simulations. The average value of all the simulations is: $0.500\pm0.004$.

In summary, if we compare these artificial loops with the ones created in the network, the difference in behaviour is clear.  Even though network loops analysed in this work are generally larger than our artificial loops, they decay much faster (with respect to their length), i.e., the lifetime of network loops exhibit a linear scaling with respect their initial length, as opposed to a quadratic lifetime of artificial loops. Besides, the velocities of network loops are lower than the ones expected from a Nambu-Goto behaviour. Artificial loops seem to 
follow closely the Nambu-Goto behaviour in velocity, though they slowly lose energy.

\section{A model for loop decay}
\label{s:DecMod}

In this section we outline a model which can explain the decay of string loops by the non-linear interaction of 
degrees of freedom living on the string with massive radiation in the full space. 
Such modes are known to exist: they are the dilatational or ``sausage'' modes of a Nielsen-Olesen vortex \cite{Goodband:1995rt}, 
and they are also known to couple to massive radiation \cite{Arodz:1995pt,Arodz:1996eb}. 
At Bogomol'nyi coupling, the mass of dilatational modes is approximately 0.75 of the 
mass of the propagating modes \cite{Goodband:1995rt}.

We suppose that the extra degrees of freedom on 
the string contribute energy per unit length $e$ and 
pressure $p$, such that the mass per unit length $\mu$ and tension $T$ are
\ben
\mu = \mu_0 + e, \quad T = \mu_0 - p ,
\een 
where $\mu_0$ is the energy per unit length of the straight static string solution. 
We suppose that the rest frame string length is $\ell$, so that the total energy is $E = \mu \ell$. 
The rate of change of the total energy is then
\ben
\dot{E} = \dot{e} \ell + \mu \dot\ell .
\een
If  these extra degrees of freedom are indeed the dilatational modes, 
then they can interact and excite massive radiation. 
It is known that the energy loss rate of a 2-dimensional vortex with the 
dilatational mode excited is proportional to the square 
of the excitation energy \cite{Arodz:1996eb}.  
This behaviour has also been   observed in numerical 
simulations of kinks in a 1+1 dimensional field theory \cite{Blanco-Pillado:2020smt}, which also 
possess dilatational modes (see e.g. \cite{Rajaraman:1982is}). 
One can therefore infer that the energy loss rate per unit length 
of a string is proportional to the square of the energy per unit length 
\ben
P_\text{rad} = - \frac{\ka}{\phi_0} e^2,
\een
where $\ka$ is a dimensionless coupling constant, and the expectation value of the field 
$\phi_0$ sets the scale.  
This is consistent with a picture of the interaction and energy loss as 
due to scattering in $1 + 1$ dimensions. 

Hence the rate of loss of the total energy is 
\ben
\dot{E} = - \frac{\ka}{\phi_0} e^2\ell .
\een
Some fraction $f$ of the energy will be lost from the extra degrees of freedom and 
therefore $(1-f)$ from the string itself.
The energy per unit length in the extra degrees of freedom 
can also change due to the change in the string background on which 
the modes propagate.  From energy-momentum conservation on the string world sheet 
one can estimate 
\ben
\dot e \simeq - \frac{\dot\ell}{\ell} (e+p) ,
\een
and so one can write 
\ben
\dot{e} \simeq - (e+p) \frac{\dot\ell}{\ell}  - f \frac{\ka}{\phi_0} e^2.
\een
Hence, 
\ben
(\mu_0 - p) \dot\ell \simeq -  (1-f)\frac{\ka}{\phi_0} e^2 \ell .
\een

\subsection{Lifetime proportional to length squared}
\begin{figure}[h]
    \centering
    \includegraphics[width=\columnwidth]{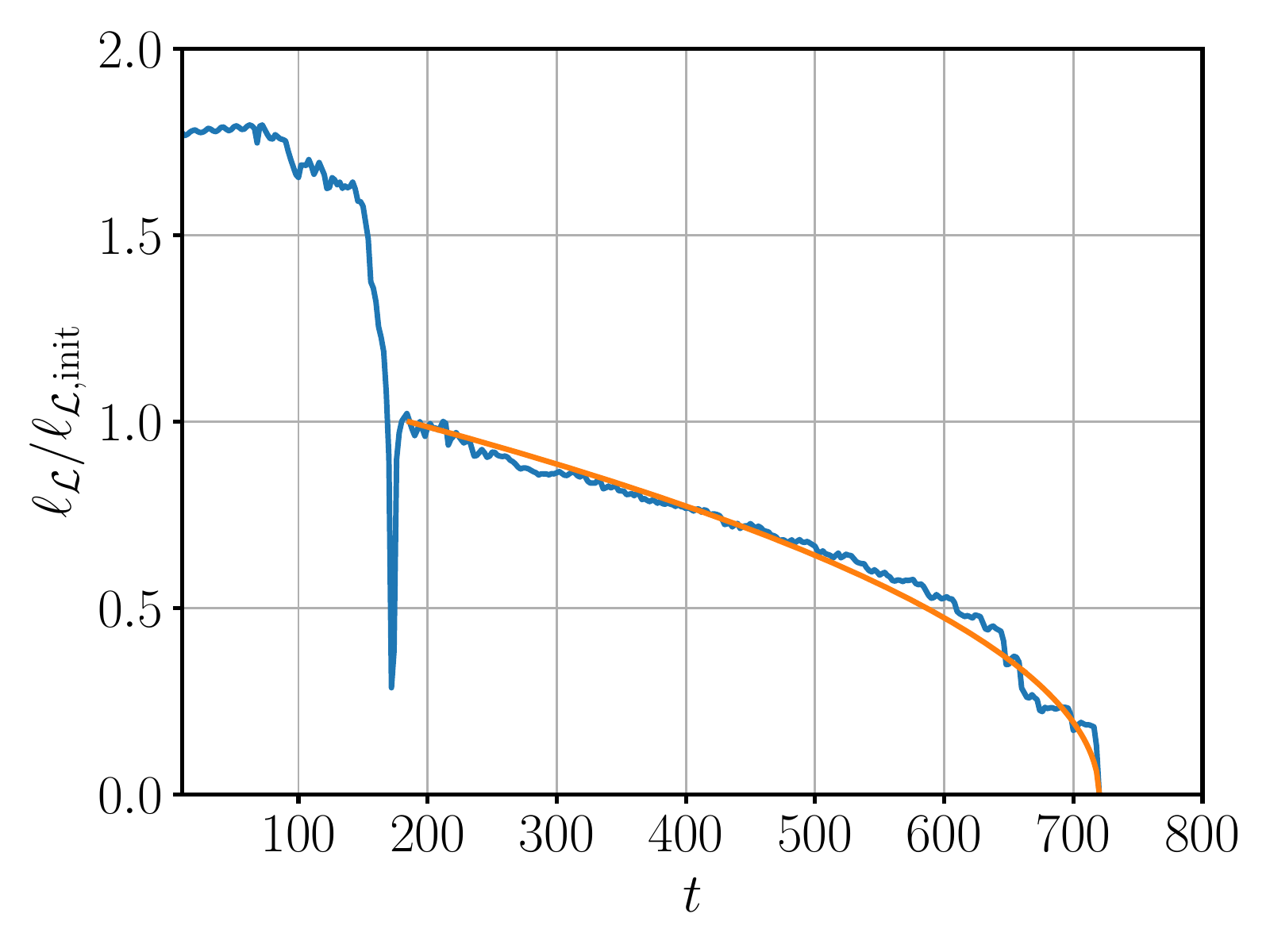}
    \caption{\label{ell_lag_decay_simulation_vs_model_outer_fit}
Comparison between the length decay obtained from the simulation of an artificial loop 
(the same as in Fig.~\ref{artifvel}) 
and theoretical prediction from Eq.~(\ref{e:DecLaw}). 
We approximate the string configuration by a single loop decaying when the second 
loop in the simulation disappears.}
 \end{figure}

Let us first suppose that energy in the extra degrees of freedom is 
much less than that of the string background, $ e \ll \mu_0$, and $p \ll \mu_0$.
Then we can neglect the fraction $f$, 
and the invariant length of the string decays as 
\ben
\dot\ell = -\frac{\ka}{\phi_0^3} e^2 \ell .
\een
We also suppose that the equation of state of the extra degrees of freedom is $p=0$, 
consistent with their being non-relativistic massive modes. 
Hence the energy per unit length of the extra degrees of freedom 
increases with decreasing $\ell$, as 
\ben
e = e_0 (\ell_0/\ell) .
\een
We can then write 
\ben
\dot\ell = -\frac{\ka e_0^2 }{\phi_0^3} \left( \frac{\ell_0}{\ell} \right)^2 \ell ,
\een
which has the solution
\ben
\ell^2 = \ell_0^2 \left (1  -  \frac{2\ka e_0^2 }{\phi_0^3} t\right) .
\een
We then suppose that the initial energy in the degrees of freedom is 
associated with the initial curvature of the string, so 
assuming that the string curvature is of order $\ell^{-1}$,
\ben
\label{e:IniEne}
e_0 = \ep \phi_0/\ellInit, 
\een
we can solve the equation for $\ell$, obtaining
\ben
\ell^2 = \ell_0^2 \left (1  -  \frac{2\ka \ep^2 }{\phi_0 \ellInit^2} t\right)\,.
\een
Thus the lifetime of the loop is 
\ben
t_\text{life} = \frac{\phi_0 \ellInit^2}{2\ka \ep^2 } ,
\een
which is proportional to the square of the length, as observed.  
One can write the decay of the length in the form 
\ben
\label{e:DecLaw}
\frac{\ell}{\ellInit} = \left (1  -  \frac{t-t_{\textrm{init}} }{t_\text{life}} \right)^\half .
\een
This prediction is compared with a simulation of artificial loops in Fig.~\ref{ell_lag_decay_simulation_vs_model_outer_fit}.  
In view of the simple approximations, and the fact that the string is 
contained in two loops, the fit is good. 

The decay law (\ref{e:DecLaw}) generalises the model of Ref.~\cite{Matsunami:2019fss}, which 
models the energy loss as due to the collision of kinks.  
A kink is just one kind of perturbation raising the energy of the string;
a single kink can be expected to increase the energy of the string by $\delta E \sim \phi_0$, 
and so with $n_\text{k}$ kinks, the string has initial extra energy 
per unit length $e_0 \sim n_\text{k} \phi_0/\ellInit$, scaling in the same way 
as Eq.~(\ref{e:IniEne}).  
Hence the decay law (\ref{e:DecLaw}) applies to kinky strings as well, at 
least when averaged over many kink collisions.

\subsection{Lifetime proportional to length}

We now suppose that there are two components to the extra degrees of freedom, $e_a$ and $e_b$. 
They could, for example, be short-wavelength and long-wavelength modes.

We assume that one of these is excited by the curvature of the string, as before, 
but that the other is saturated at a
constant value $e$, which is of order $\mu_0$. 
This will lead to a reduction in the velocity of waves on the string $\cs$ to 
\ben
\cs^2 = 1/(1 + e/\mu_0) .
\een
Then 
\ben
e_a = \ep_a \phi_0/\ell, \quad e_b = \ep_b \mu_0 .
\een
Staying with the assumption that the pressure of the 
extra degrees of freedom is negligible, the decay of string length then goes as 
\ben
\mu_0\dot\ell = - (1 - f) \frac{\ka}{\phi_0} e_a e_b \ell = -   (1 - f)\ka \ep_a \ep_b \mu_0 .
\een
If we assume that the fraction of the energy lost by the extra degrees of freedom is 
constant, we arrive at a decay law linear in time,
\ben
\dot\ell = -   \tilde{\ka} {\ep_a \ep_b} ,
\een
where we have absorbed $1-f$ into the constant $ \tilde{\ka} $. 
There is evidence for a saturated $e_b$, as the mean square velocity  of 
the network loops is 
observed to be around 0.4, less than the Minkowski spacetime value of 0.5 for a NG string.
This indicates that the square of the propagation speed of 
waves on the string is around $0.8$, and hence 
$e/\mu_0 \simeq 0.25$. 

There is a significant variability in the decay rates of individual loops, which presumably reflects 
the way that the initial conditions perturb the strings, but the saturation of the energy of the 
extra degrees of freedom seems to be generic. That this saturation always seems to occur 
for network loops points towards the presence an instability, whose nature deserves further investigation.

\section{Conclusions}
\label{sec:conclusions}
In this paper we have investigated the dynamics of Abelian Higgs loops, focusing on their lifetime  and their decay mechanisms. This characterisation has deep implication on the properties of networks of cosmic strings, in particular, both the cosmic ray and 
gravitational wave power of such networks depends strongly on the properties of the loops.  The GW predictions and bounds obtained for cosmic strings are generally obtained by studying the Nambu-Goto dynamics of strings, and the question lies whether string loops in a field theory follow such dynamics, or they have different decay channels.

We have studied both the length and the velocity of the loops,  created in two different ways. On the one hand, we have formed a network of strings from 
cooled random fields, and followed the dynamics of loops created by network evolution. 
On the other, we have set up initial conditions carefully tuned to create long-lived oscillating loops, which we term ``artificial'' loops.

For the randomly created loops,  we have simulated about 100 different cases (with different random seeds, initial correlation lengths and lattice resolutions). Only about one third of those cases resulted in loops created by intersection of other loops, which we consider to be more representative of the typical loop created by a scaling string network. We term these ``network'' loops. 
About half of the simulations ended strings wound around the lattice, which had periodic boundary conditions. In about one sixth the loops present in the initial state evaporated without intersection.

All randomly created loops decayed rapidly, shrinking either to zero size or 
to strings wrapping the simulation volume within a time of order $\ellInit$, their initial length. 
The lifetime of network loops, as defined above, is proportional to $\ellInit$, 
with a proportionality constant of $0.14\pm0.04$. 
The lifetimes show no tendency to increase with the initial length. 
These results are consistent with scaling network simulations \cite{Hindmarsh:2017qff} and with previous studies of decaying loops \cite{Hindmarsh:2008dw}, and extend them to loop sizes up to around 6000 inverse mass units.

 The mean square velocity for the network loops is observed to be $0.40 \pm 0.04$, 
lower than the value $0.5$ for a loop following Nambu-Goto dynamics in Minkowski spacetime. 

Our artificial loops were created by the collision of two pairs of wrapping strings with either a sawtooth or sinusoidal initial shape. This set of loops 
behaved quite differently:  their lifetime was proportional to the square of their initial length, in line with the results of \cite{Matsunami:2019fss}, who studied artificial loops created by the collision of moving straight strings. The mean square velocity of these loops was $ 0.500 \pm 0.004$, very close to the prediction of Nambu-Goto dynamics.
The collision and reconnection of the strings resulted in the loss of about half the energy in the initial state, independently of the ratio of the string length to the string width, 
which is contrary to the ``traditional'' picture, where energy is lost only from a few inverse mass units around the site of the self-intersection. 

The fact that we are able to simulate loops obeying approximate Nambu-Goto dynamics and loops which rapidly lose energy by radiation, 
leads us to conclude there is an intrinsic difference between network loops and artificial loops.

We propose a model to explain the difference, in terms of the excitation of degrees of freedom on the string, which could be 
dilatational modes.  If the degrees of freedom are only weakly excited, the decay rate of loop length is inversely proportional to the length, 
leading to a quadratic dependence of the lifetime on the initial length. 
If the modes are strongly excited, with energy per unit length comparable with that of the background string solution, 
the decay rate of loop length is independent of the length, and the loop lifetime depends linearly on the initial length. 
There is evidence for excitations with energy of order 20\% of the background string from the lower mean square velocity of network loops. 
The fact that the extra degrees of freedom are consistently excited by the collapse of initially smooth field configurations  
indicates the presence of an intrinsic instability in the background of an evolving string.

The nature of this instability is somewhat mysterious. 
Some clues are available from the available simulations, in that if loops are somehow prevented from collapsing, 
either by spinning (in the artificial loop case) or by wrapping, the instability seems to shut off. 
We also note that a massive mode propagating in the effective 1+1 dimensional spacetime of an oscillating string 
can undergo parametric resonance. A dedicated study is necessary to investigate. 
As radiating dilatational modes can also be excited on 1+1 dimensional kinks \cite{Blanco-Pillado:2020smt}, 
the same effect may also be found in 2+1 dimensional simulations of a real scalar field theory with 
a spontaneously broken $Z_2$ symmetry, for which simulations will be less demanding.

It may be the case that most of the network loops behave as the loops found in this work, 
but there could be a fraction of them that are well-described by Nambu-Goto dynamics, a fraction which is 
sufficiently small that we were not lucky enough to create one. 
To accommodate this uncertainty in modelling of observational signals of strings, we propose a parameter $\fNG$, 
 the fraction of loops  in a network following Nambu-Goto dynamics, which can be used to quantify and recalculate 
 the bounds on cosmic strings from cosmic rays or gravitational waves. 
 Using the ``rule of three'' statistics on our sample, $\fNG< 0.1$ at 95\% confidence level, a bound which 
 includes zero.  
 
There are significant physical implications of the scarcity of Nambu-Goto-like loops in a field theory string network.  
For example,  if the recent NANOgrav report of a possible stochastic gravitational wave background 
at frequencies around $1$ yr$^{-1}$ \cite{Arzoumanian:2020vkk} is due to cosmic strings \cite{Ellis:2020ena,Blasi:2020mfx} with string tension $\gmu \sim 10^{-10}$, 
their fields cannot couple significantly to the Standard Model, 
as there would be too much gamma ray production to be consistent with observation 
\cite{Mota:2014uka}. If the strings are almost completely decoupled from the 
Standard Model, they could have larger string tension, 
in the range $10^{-10}\fNG^{-1} \lesssim \gmu \lesssim 10^{-7} $.
The upper bound on $\gmu$ comes from the CMB \cite{Lizarraga:2016onn}, and is purely gravitational. 
Hence $\fNG \gtrsim 10^{-3}$ is required for hidden sector field theory strings to 
account for the NANOgrav signal.

\begin{acknowledgments}
MH (ORCID ID 0000-0002-9307-437X) acknowledges support from 
the Science and Technology Facilities Council (grant number ST/L000504/1) and 
the Academy of Finland (project 333609). 
JL (ORCID ID 0000-0002-1198-3191), AU (ORCID ID 0000-0002-0238-8390) and JU (ORCID ID 0000-0002-4221-2859) acknowledge support from Eusko Jaurlaritza (IT-979-16) and PGC2018-094626-B-C21 (MCIU/AEl/FEDER,UE). The work has been performed under the Project HPC-EUROPA3 (INFRAIA-2016-1-730897), with the support of the EC Research Innovation Action under the H2020 Programme; in particular, AU gratefully acknowledges the support of Department of Physics of the University of Helsinki and the computer resources and technical support provided by CSC Finland, as well as support from Eusko Jaurlaritza grant IKASIKER.
\end{acknowledgments}


\vspace{2cm}
\appendix
\section{Resolution tests}
\label{app:A}

We have performed a series of tests to check for the effect of lattice spacing on the decay of string loops in our simulations. In the case of the loops formed from the evolution of a network, we compare loops by coarse-graining the field configuration of 
the simulation with the finest lattice spacing.
In the case of artificial loops the strings are automatically prepared around the ``seed'' curve in a field configuration 
appropriate to the resolution, and coarse-graining is not needed.

\subsection{Networks}

We aim to establish the effect of the lattice-spacing $\dx$ on the lifetime of a loop created during the evolution of a network.
We run a simulation which ends with an evaporating loop using the finest resolution $\dx=0.125$ in a box with $N=1024$ lattice-points per side. 
Then, visualising the simulation box,  we establish the instant at which the final string loops are formed. We then rerun the simulation, and store the field values for the whole simulation box at 10 time-units before the loop is formed.
The field values are then used to create the initial conditions of the  boxes with lattice spacings $\dx=0.25$ and
$\dx=0.5$, and points $N=512$ and $N=256$, by averaging over adjacent sites \cite{Hindmarsh:2017qff}.

\begin{figure}[h]
    \centering
    \includegraphics[width=\columnwidth]{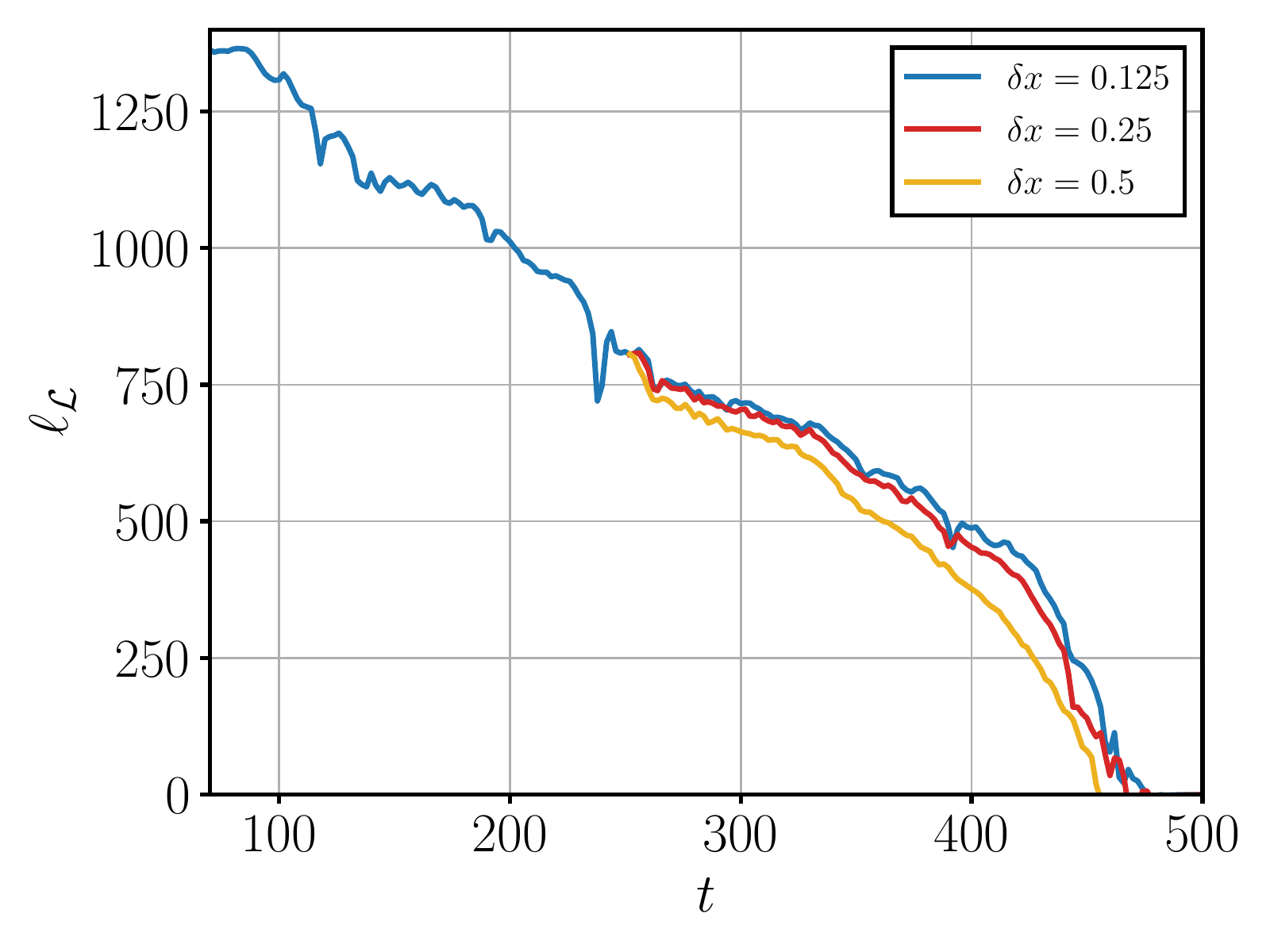}
    \caption{\label{fig:reso_ell_lag_network_general}
    Evolution of the loop length estimator for correlation length $\cl=25$. Each colour represents a different resolution, yellow for $\dx=0.5$, red for $\dx=0.25$ and blue for $\dx=0.125$.}
 \end{figure}

\begin{figure}[h]
    \centering
     \includegraphics[width=\columnwidth]{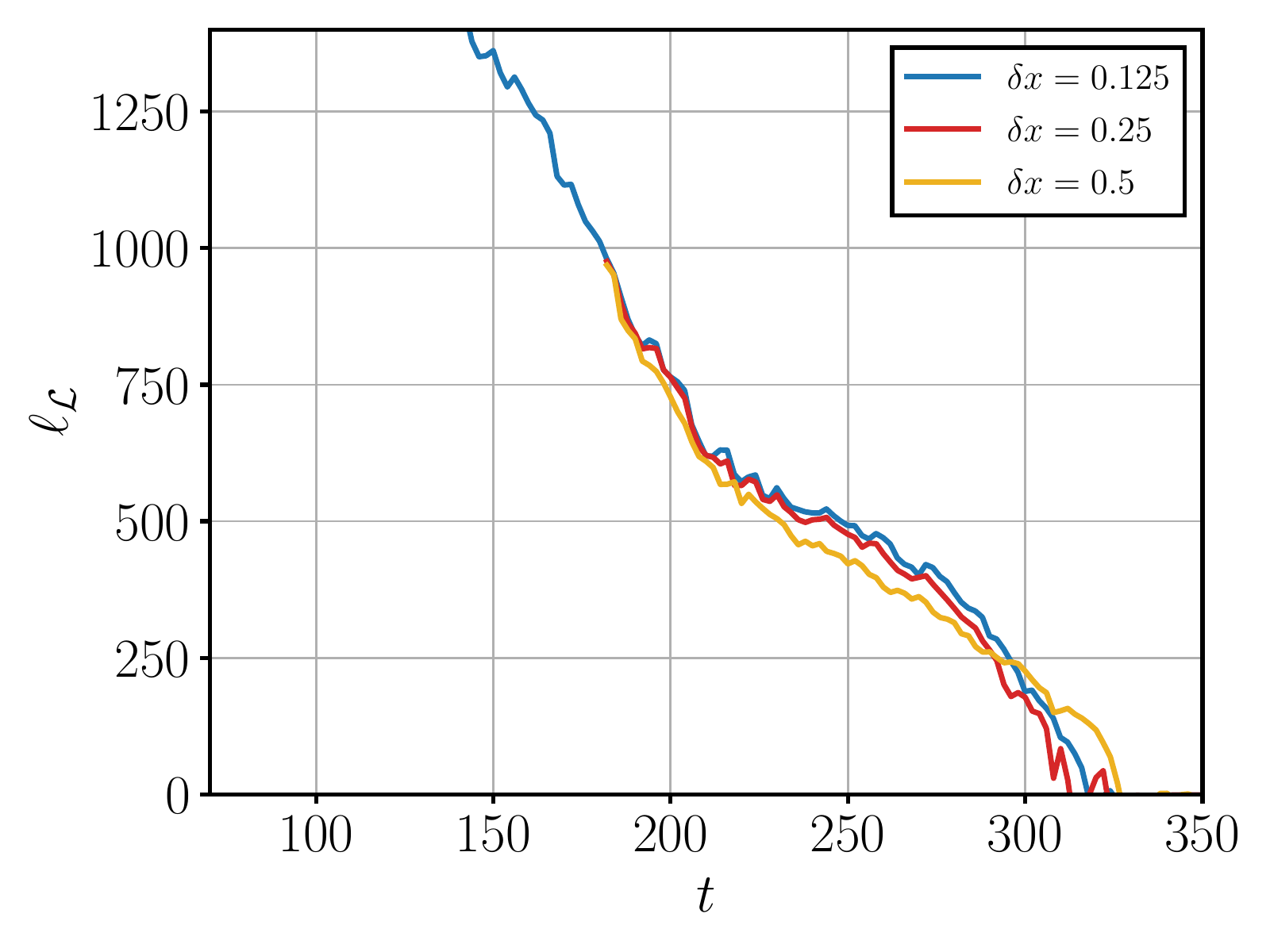}
    \caption{\label{fig:reso_ell_lag_network_split}
    Evolution of the loop length estimator for correlation length $\cl=15$. Each colour represents a different resolution, yellow for $\dx=0.5$, red for $\dx=0.25$ and blue for $\dx=0.125$.}
 \end{figure}

We run these simulations and compare the outcomes, as shown in Figs.~\ref{fig:reso_ell_lag_network_general} and~\ref{fig:reso_ell_lag_network_split}. 

Fig.~\ref{fig:reso_ell_lag_network_general} shows the evolution of the string network length $\Llag$ for a simulation with $\cl=25$. Each colour represents a different lattice resolution, blue for $\dx=0.125$, red for $\dx=0.25$ and yellow for $\dx=0.5$. As explained before, the initial field configurations of $\dx=0.5$ and $\dx=0.25$ are obtained at the moment at which the loop is created for $\dx=0.125$, and so their lines start later (in this case at  $t\approx 250$). 
It shows the general behaviour we have seen for most of the resolution tests performed.   It can be seen how the lifetime of the loop decreases when the resolution gets lower, which means a faster decay of loops if the resolution is reduced. For $\dx=0.25$ this difference with $\dx=0.125$ exists but it is not substantial. For $\dx=0.5$ the differences become somewhat bigger, 
 but is still within around 5\% of the lifetime at $\dx=0.125$.

Fig.~\ref{fig:reso_ell_lag_network_split} shows another simulation, with $\cl=15$, and where the loop is formed at  $t\approx 180$. The colours correspond to the same cases as in Fig.~\ref{fig:reso_ell_lag_network_general}. This figure also shows  some differences in the length decay of the resolution $\dx=0.5$. In the early stages of the evolution, the loop evolved at the lowest resolution  loses length more rapidly, as expected. However, at around $t\approx 270$, this behaviour suddenly changes, and the loop in the simulation with $\dx=0.5$ becomes the longest lasting loop. This behaviour is not typical.

 \begin{figure}[t]
    \centering
    \includegraphics[width=0.68\columnwidth]{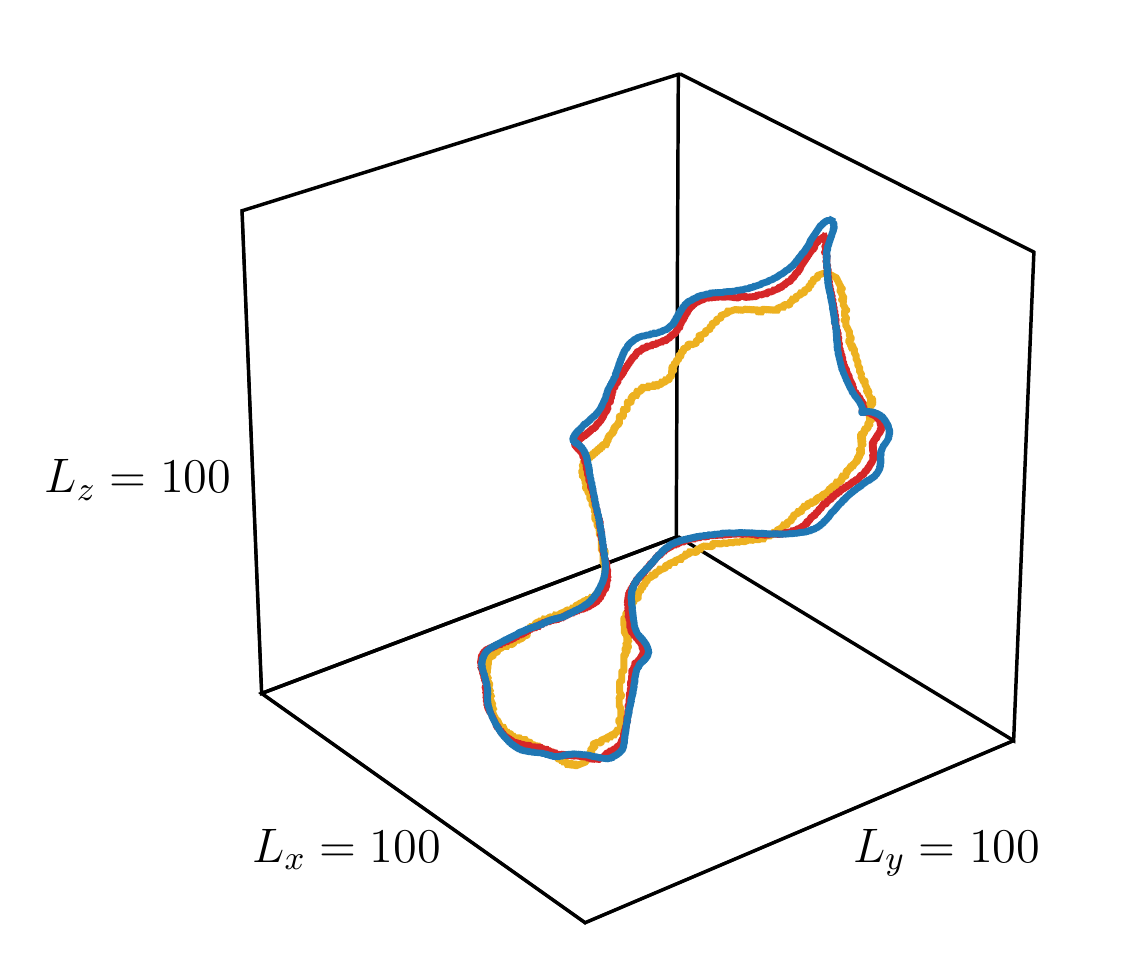}\\
    \includegraphics[width=0.68\columnwidth]{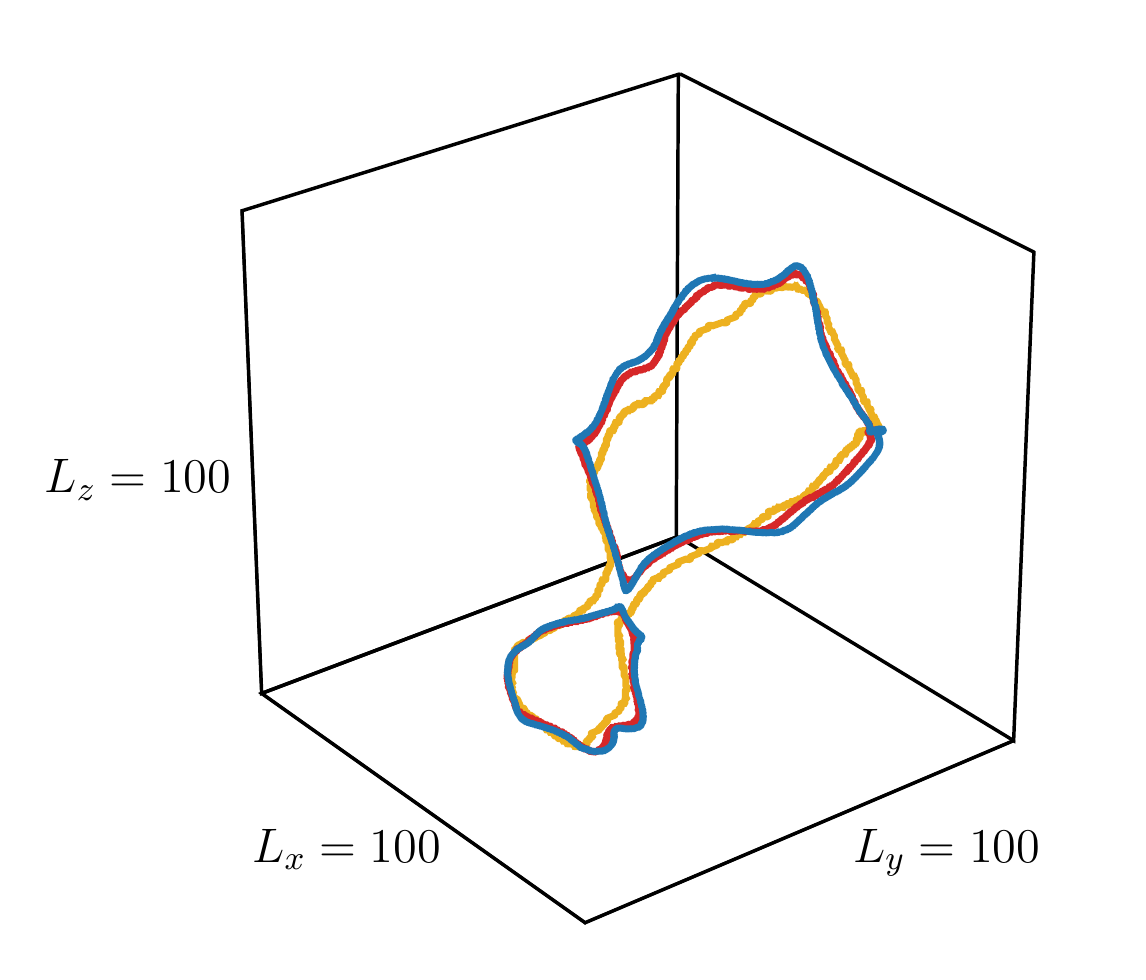}\\
    \includegraphics[width=0.68\columnwidth]{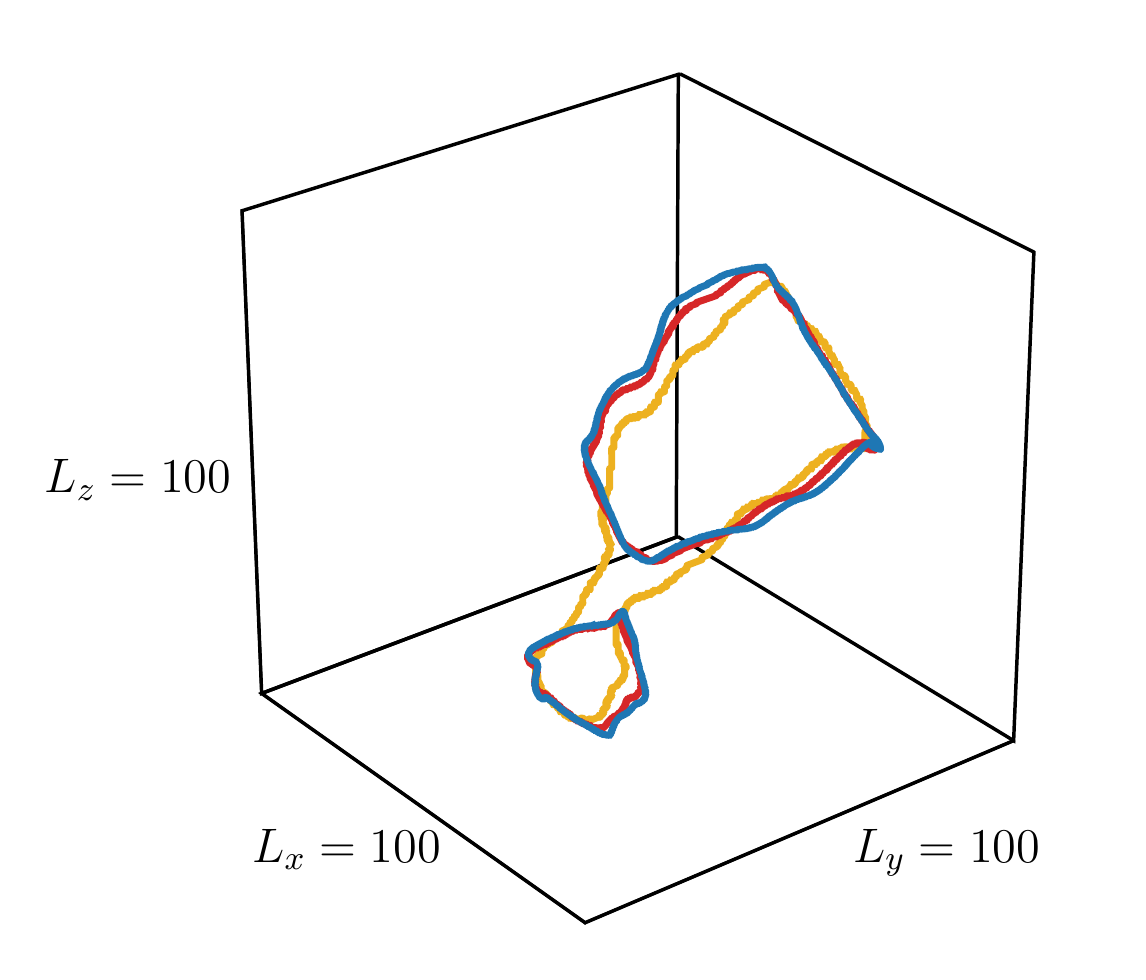}\\
    \caption{\label{fig:reso_split}
   Sequence of snapshots corresponding to the change of behaviour in Fig.~\ref{fig:reso_ell_lag_network_general} around $t\approx 270$, showing how the loop fragmentation is avoided at $\dx = 0.5$. Starting from the top each snapshot corresponds to $t=262$, $268$ and $274$. The colours represent, once again, different lattice spacing: blue for $\dx=0.125$, red for $\dx=0.25$ and yellow for $\dx=0.5$}
 \end{figure}

Fig.~\ref{fig:reso_split} shows a sequence of snapshots that illustrate why this abrupt change happens on the decay.  From top to bottom, with time units $t=262$, $268$ and $274$ respectively, we can see how for resolutions $\dx=0.125$ and $\dx=0.25$ the loop splits in two. However, this fragmentation process does not happen for $\dx=0.5$, and therefore it takes more time for the loop to fully decay.

In summary, the resolution tests that we have performed indicate similar evolution of the length estimators. The lifetime of the loops obtained from all different resolutions are within $5\%$. We decided to be somewhat conservative, and taking into account our numerical resources, opted to perform simulations with $\dx=0.25$ and $\dx=0.125$.

\subsection{Artificial strings}

The study of the loops created by the mechanism described in Sec.~\ref{subsec:AI} also requires 
resolution checks. In this case there is no need to coarse-grain, since we control the formation of the loop. 

\begin{figure}[h]
    \centering
    \includegraphics[width=\columnwidth]{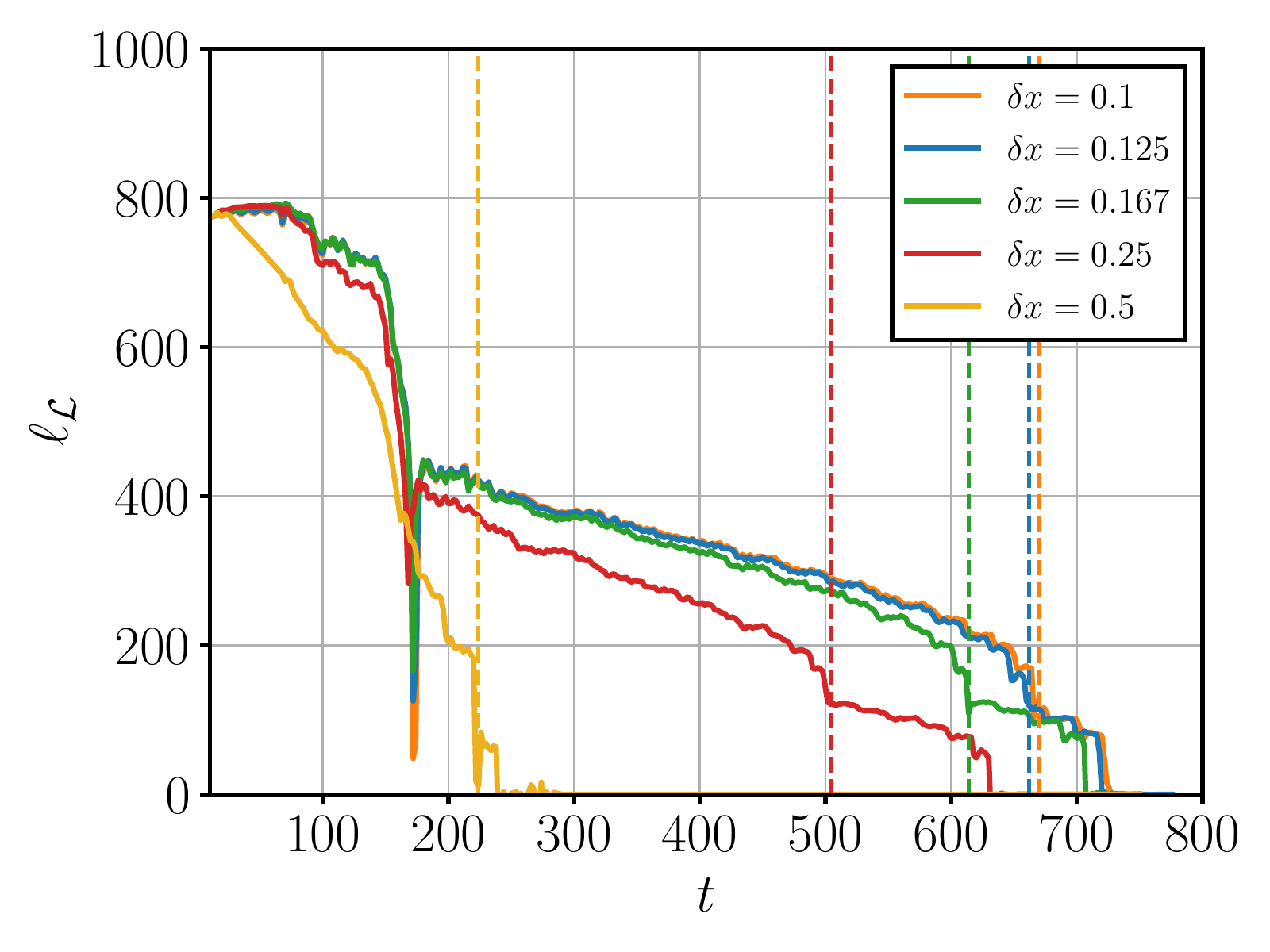}
    \caption{\label{fig:apparti_res1}
    Evolution of the loop length estimator $\Llag$ for different resolutions. Dashed vertical lines represent the moment where the inner loop disappears. Each colour represents a different resolution, yellow for $\dx=0.5$, red for $\dx=0.25$, green for $\dx=0.167$, blue for $\dx=0.125$ and orange for $\dx=0.1$.}
 \end{figure}

Fig.~\ref{fig:apparti_res1} shows the evolution of the length estimator for the different resolutions we have tested. Each colour refers to a different resolution and lattice size (ensuring all configurations are equivalent), yellow for $(N=256,\dx=0.5)$, red for $(N=512,\dx=0.25)$, green for $(N=768,\dx=0.167)$, blue for $(N=1024,\dx=0.125)$ and orange for $(N=1280,\dx=0.1)$. The dashed vertical line is used to show the moment at which the inner loop disappears.

This figure shows that the length decreases faster for $\dx=0.5$ than for the other resolutions. As explained in the main test, we aim at configurations that avoided a double-line collapse, but 
this resolution is too low to avoid it and the loop decays without oscillating. Resolutions  of $\dx=0.25$ and higher succeed in avoiding the double-line collapse.

However, comparing the resolutions which avoid the collapse we observe that the loops in resolutions  $\dx=0.25$  and $\dx=0.167$ decay faster than in higher resolutions, whereas $\dx=0.125$ and  $\dx=0.1$  the decay is practically identical. In this case, the collapse of the inner loops only differs by  8 time steps.

 \begin{figure}[h]
    \centering
    \includegraphics[width=\columnwidth]{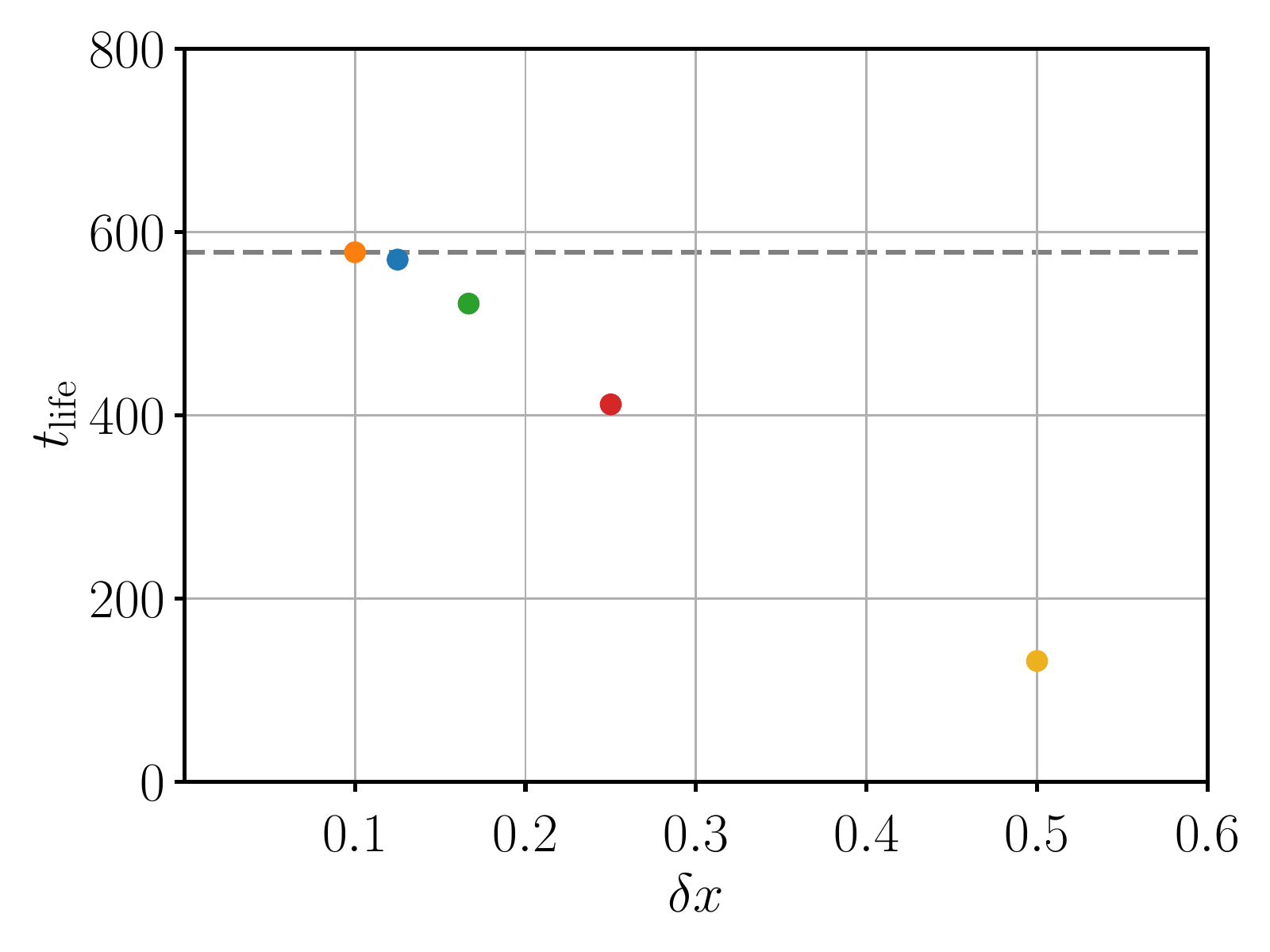}
    \caption{\label{fig:apparti_res2}
    Lifetime of the inner loops versus spatial resolution for artificially created loops.}
 \end{figure}

The convergence can be seen in Fig.~\ref{fig:apparti_res2}, which shows the lifetime of the inner loop for each resolution. The dashed horizontal grey line refers to the lifetime of the finest resolution, $\dx=0.1$, we have tested. The lifetime increases as the resolution increases, until we observe that the lifetime of the loop at resolution $\dx=0.125$ is quite similar to
that at $\dx=0.1$.

In view of these results, and our numerical resources, we decided to perform simulations with  $\dx=0.125$.

\bibliography{CosmicStrings}

\begin{thebibliography}{45}
\bibitem [{AHL()}]{AHLoopVideos}%
  \BibitemOpen
  \href@noop {} {}\bibinfo {howpublished} {Videos can be found as ancillary
  files in the arxiv submission.}\BibitemShut {Stop}%
\end{thebibliography}



\end{document}